\documentclass[twocolumn,amssymb,amsmath]{revtex4}
\usepackage{CJK}
\pdfoutput=1
\usepackage{graphicx}

\begin{document}
\title{Magnon peak lineshape in the transverse dynamical structure factor of a ferromagnetically polarized easy-axis $XXZ$ chain at low temperatures}
\author{P.~N.~Bibikov}
\affiliation{\it Russian State Hydrometeorological University, Saint-Petersburg, Russia}

\begin{abstract}

The ferromagnetically polarized gapped XXZ spin chain is
studied at low temperatures. Utilizing only the one- and two-magnon spectrums
and focusing on the magnon-creation contribution to the transverse dynamical susceptibility, we represent the latter in the form of the Dyson equation. Then, following the well known correspondence between the imaginary part of magnetic susceptibility and dynamical structure factor, we get the low-temperature formula for the magnon-peak lineshape. The suggested approach is effective only if the processes related to magnon creations and to
transitions from magnons to coupled magnon pairs are energetically separated. As it is shown in the paper, such separation is inherent in
the easy-axis chains with rather strong anisotropy.
We present several plots and discuss their lineshapes. As the supplemental result we obtain integral representations for the temperature-dependent magnon resonance shift and
the parameter which is usually associated with the decay rate. The low-temperature behavior of the resonance shift is studied in details. All calculations are performed up to {\it controllable} error $o({\rm e}^{-\beta E_{gap}})$.
\end{abstract}

\maketitle

\section{Introduction}

Contrary to the zero-temperature case \cite{1}, systematic study of $T>0$ correlations in spin chains was developed only in the present century \cite{2,3,4,5,6,7,8,9,10}. Among the {\it dynamical} correlation functions, very important is the (measurable by neutron scattering \cite{11}) transverse dynamical structure factor (TDSF). In the present paper we shall study its low-temperature asymptotics in the ferromagnetic phase for the
XXZ spin chain related to the Hamiltonian
\begin{eqnarray}\label{ham}
&&\hat H^{(XXZ)}=-\sum_{n=1}^N\Big[\frac{J_{\bot}}{2}\Big({\bf S}^+_n{\bf S}^-_{n+1}+{\bf S}^-_n{\bf S}^+_{n+1}\Big)\nonumber\\
&&+J_z\Big({\bf S}^z_n{\bf S}^z_{n+1}-\frac{1}{4}\Big)+h\Big({\bf S}_n^z-\frac{1}{2}\Big)\Big],\qquad h\geq0,
\end{eqnarray}
where ${\bf S}_n^z$ and ${\bf S}_n^{\pm}={\bf S}_n^x\pm i{\bf S}_n^y$ ($n=1,\dots,N$) are the usual spin-1/2 operators attached to the chain sites.
Hamiltonian \eqref{ham} acts on the tensor product of $N$ two-dimensional vector spaces spanned on
spin-up and spin-down states $|\uparrow\rangle$ and $|\downarrow\rangle$.
The final result will be obtained in the thermodynamical limit $N=\infty$; however, initially we begin with the periodical model, supposing that
\begin{equation}\label{period}
{\bf S}_{N+1}\equiv{\bf S}_1.
\end{equation}

We shall treat the model \eqref{ham} only under the condition
\begin{equation}\label{addcond}
E_{\rm gap}=h+J_z-|J_{\bot}|>0,
\end{equation}
[as it follows from \eqref{emagn} $E_{\rm gap}$ is the energy gap between the ground state and the one-magnon sector] under which the system is gapped and has the ferromagnetically polarized, zero-energy ground state,
\begin{equation}\label{ground}
|\emptyset\rangle=|\uparrow_1\rangle\otimes\dots\otimes|\uparrow_N\rangle,\qquad\hat H|\emptyset\rangle=0.
\end{equation}

TDSF may be defined by two equivalent ways. The former is the spectral decomposition
\begin{equation}\label{tdsfspec}
S(\omega,q,T)=\sum_{\mu,\nu}\frac{{\rm e}^{-\beta E_{\nu}}|\langle\nu|{\bf S}^+(q)|\mu\rangle|^2\delta(\omega+E_{\nu}-E_{\mu})}{Z(T)},
\end{equation}
where $\mu$ and $\nu$ enumerate the Hamiltonian eigenstates and $Z(T)\equiv{\rm tr}({\rm e}^{-\beta\hat H})$.
The next definition,
\begin{equation}\label{tdsf}
S(\omega,q,T)=-\frac{{\rm Im}\chi(\omega,q,T)}{\pi(1-{\rm e}^{-\beta\omega})},\qquad\omega\neq0,
\end{equation}
is based on the correspondence between TDSF and the transverse dynamical magnetic susceptibility
\begin{equation}\label{susc}
\chi(\omega,q,T)=\langle\langle{\bf S}^+(q),{\bf S}^-(-q)\rangle\rangle.
\end{equation}
Here
\begin{equation}\label{s(q)}
{\bf S}(q)\equiv\frac{1}{\sqrt{N}}\sum_{n=1}^N{\rm e}^{-iqn}{\bf S}_n,\qquad{\rm e}^{iqN}=1,
\end{equation}
and for an arbitrary pair of operators ${\cal A}$ and ${\cal B}$ there are two equivalent definitions
of the real two-time commutator retarded one-magnon Green function $\langle\langle{\cal A},{\cal B}\rangle\rangle$
\cite{12}
\begin{subequations}\label{twodefs}
\begin{eqnarray}
&&\langle\langle{\cal A},{\cal B}\rangle\rangle\equiv\frac{1}{i}\int_0^{\infty}dt{\rm e}^{i(\omega+i\epsilon)t}\langle[{\cal A}(t),{\cal B}]\rangle,\\
&&\langle\langle{\cal A},{\cal B}\rangle\rangle\equiv\frac{1}{i}\int_0^{\infty}dt{\rm e}^{i(\omega+i\epsilon)t}\langle[{\cal A},{\cal B}(-t)]\rangle.
\end{eqnarray}
\end{subequations}
Here ${\cal A}(t)={\rm e}^{iHt}{\cal A}{\rm e}^{-iHt}$ and $\langle{\cal A}\rangle\equiv{\rm tr}({\rm e}^{-\beta\hat H}{\cal A})/Z(T)$.

Since [under condition \eqref{addcond}] the system is gapped, it is a temptation to evaluate TDSF directly by formula \eqref{tdsfspec} and,  using the machinery of low-temperature  formfactor expansions \cite{6,7,8,9}, reduce \eqref{tdsfspec} to the form
\begin{equation}\label{Sclust}
S(\omega,q,T)=S_0(\omega,q)+\sum_{m=1}^{\infty}S_m(\omega,q,T),
\end{equation}
where each $S_m(\omega,q,T)$ depends on matrix elements $\langle\nu|{\bf S}^+(q)|\mu\rangle$ between $j$-magnon states $\langle\nu|$ and
$j+1$-magnon states $|\mu\rangle$ with $j\leq m$, so that
\begin{equation}
S_m(\omega,q,T)=O(\zeta_{\rm g}^m),\qquad\zeta_{\rm g}\equiv{\rm e}^{-\beta E_{\rm gap}}.
\end{equation}

At first glance, it seems natural that even a few number of terms in the expansion \eqref{Sclust} may ensure a good approximation for TDSF
in the low-temperature regime,
\begin{equation}\label{lowT}
{\rm e}^{-\beta E_{\rm gap}}\ll1,\qquad\beta\equiv\frac{1}{k_BT},
\end{equation}
so that the expression for $S(\omega,q,T)$ will be governed by the low-lying spectrum.
It is well known, however, that {\it in the ferromagnetic phase} \eqref{addcond} and \eqref{ground}, when
\begin{equation}\label{S0}
S_0(\omega,q)=\delta[\omega-E_{\rm magn}(q)],
\end{equation}
this approach fails due to the $T=0$ delta-singularity which cannot be canceled by any finite number of terms in the sum \eqref{Sclust}.

In order to avoid this pathology and get a broadened $T>0$ lineshape remaining in the framework of power expansions, it was suggested in Refs. \cite{6,7,8,9} to utilize the formulas \eqref{tdsf} and \eqref{susc} instead of \eqref{tdsfspec}. Since the direct application of the $\rm K\ddot all\acute e n$-Lehmann spectral decomposition \cite{12}
\begin{equation}\label{lehm}
\langle\langle{\cal A},{\cal B}\rangle\rangle=\sum_{\mu,\nu}\frac{({\rm e}^{-\beta E_{\nu}}-{\rm e}^{-\beta E_{\mu}})\langle\nu|{\cal A}|\mu\rangle\langle\mu|{\cal B}|\nu\rangle}{Z(T,N)(\omega+E_{\nu}-E_{\mu}+i\epsilon)},
\end{equation}
still results in the delta-singularity \eqref{S0}, it was additionally suggested to initially represent $\chi(\omega,q,T)$ in the form of the Dyson equation and obtain from \eqref{lehm} the power series for the mass operator whose imaginary part will remove the singularity and broaden the contour.

This revolutionary approach has, however, some lacks. First, it is suitable only for working with imaginary-time Matsubara Green functions, because only for them the Dyson equation was derived long ago within the rather special perturbative expansion and for rather special models (see references in Ref. \cite{13}). Next, the approach \cite{6,7,8,9} is heuristical and grounds only on the very probable conjecture that the result should be right. Hence a correct derivation of the Dyson equation still remains a {\it challenge}.

For the real-time Green functions the {\it essential progress} in this direction has been achieved by N. M. Plakida \cite{14} and then developed by Yu. A. Tserkovnikov \cite{15} (see also Ref. \cite{13}) within the alternative approach based on
the pair of equivalent equations,
\begin{subequations}\label{twoeqs}
\begin{eqnarray}
&&(\omega+i\epsilon)\langle\langle{\cal A},{\cal B}\rangle\rangle=\langle[{\cal A},{\cal B}]\rangle
+\langle\langle[{\cal A},\hat H],{\cal B}\rangle\rangle,\\
&&(\omega+i\epsilon)\langle\langle{\cal A},{\cal B}\rangle\rangle=\langle[{\cal A},{\cal B}]\rangle
+\langle\langle{\cal A},[\hat H,{\cal B}]\rangle\rangle,
\end{eqnarray}
\end{subequations}
readily following from \eqref{twodefs}. Strictly speaking, the formula presented in Ref. \cite{14} [see Eq. \eqref{dys0} in the present paper] is not yet the Dyson equation
but should be reduced to it within appropriate approximations. One of them has been suggested in Ref. \cite{10} where the gapped ferromagnetically polarized XX spin chain ($J_z=0$) in the regime \eqref{lowT} was treated. It the present paper, following this line of research, we give the {\it well-grounded} derivation of the effective low-temperature Dyson equation for the model \eqref{ham}, \eqref{addcond}, and \eqref{ground}.
The obtained result is rigorous and gives the approximation up to the {\it controllable} order $o(\zeta_{\rm g})$.

Before treating the XXZ model it is convenient to look back on the Ising ($J_{\bot}=0$) chain for which,
following \eqref{twodefs} and the rather elementary formula $\langle[{\bf S}_m^+(t),{\bf S}_n^-]\rangle=0$ $(m\neq n)$,
$\chi(\omega,q,T)$ reduces to the transverse autocorrelator $\langle\langle{\bf S}_n^+,{\bf S}_n^-\rangle\rangle$ (independently on $n$). Being exactly represented as the polar sum \cite{16}
\begin{equation}\label{Ising}
\langle\langle{\bf S}_n^+,{\bf S}_n^-\rangle\rangle=\sum_{j=1}^3\frac{A_j(T)}{\omega-\omega_j+i\epsilon},
\end{equation}
it corresponds to three different kind of processes
\begin{eqnarray}\label{proc1}
|\dots\uparrow_{n-1}\uparrow_n\uparrow_{n+1}\dots\rangle\rightarrow|\dots\uparrow_{n-1}\downarrow_n\uparrow_{n+1}\dots\rangle,\\\label{proc2}
|\dots\downarrow_{n-1}\uparrow_n\uparrow_{n+1}\dots\rangle\rightarrow|\dots\downarrow_{n-1}\downarrow_n\uparrow_{n+1}\dots\rangle,\nonumber\\
|\dots\uparrow_{n-1}\uparrow_n\downarrow_{n+1}\dots\rangle\rightarrow|\dots\uparrow_{n-1}\downarrow_n\downarrow_{n+1}\dots\rangle,\\\label{proc3}
|\dots\downarrow_{n-1}\uparrow_n\downarrow_{n+1}\dots\rangle\rightarrow|\dots\downarrow_{n-1}\downarrow_n\downarrow_{n+1}\dots\rangle.
\end{eqnarray}

Only the term, corresponding to \eqref{proc1} or, equivalently, to creations of isolated down spins (the Ising magnons) has zero activation energy (is nonzero at $T=0$).
Since all $\omega_j$ in \eqref{Ising} are different \cite{16}, the corresponding points $\omega=\omega_j$ in the $\omega$ axis for which
\begin{equation}\label{Imneq}
{\rm Im}\chi(\omega,q,T)\neq0,
\end{equation}
(the poles of $\langle\langle{\bf S}_n^+,{\bf S}_n^-\rangle\rangle$) are isolated. Hence, it is natural to suppose that at strong easy-axis anisotropy
$|\Delta|\gg1$,
\begin{equation}\label{Delta}
\Delta\equiv\frac{J_z}{J_{\bot}},
\end{equation}
the condition \eqref{Imneq} will be satisfied only inside small nonintersecting $\omega$ intervals corresponding to spreadings of the points $\{\omega_j\}$.
Among them should be the resonance magnon peak interval $[\omega_{\rm min}(q),\omega_{\rm max}(q)]$, characterizing by the property
\begin{equation}\label{emagn[]}
E_{\rm mang}(q)\in[\omega_{\rm min}(q),\omega_{\rm max}(q)].
\end{equation}
Taking
\begin{equation}\label{tdsfmagn}
S_{\rm magn}(\omega,q,T)=-\frac{{\rm Im}\chi_{\rm magn}(\omega,q,T)}{\pi(1-{\rm e}^{-\beta\omega})},\qquad\omega\neq0,
\end{equation}
where
\begin{equation}\label{separ}
\chi_{\rm magn}(\omega,q,T)\equiv\chi(\omega,q,T)\Big|_{\omega\in[\omega_{\rm min}(q),\omega_{\rm max}(q)]},
\end{equation}
one will get the magnon contribution to the TDSF.
As it will be shown in the below, up to the order $o(\zeta_{\rm g})$, $\chi_{\rm magn}(\omega,q,T)$ has the Dyson equation form \cite{10}
\begin{equation}\label{dys}
\chi_{\rm magn}(\omega,q,T)=\frac{2M(T)}{\omega-E_{\rm magn}(q)-\Sigma(\omega,q,T)},
\end{equation}
where $M(T)\equiv\langle{\bf S}^z_n\rangle$ is the average magnetization and
\begin{equation}\label{self}
\Sigma(\omega,q,T)\equiv\varepsilon(\omega,q,T)-i\Gamma(\omega,q,T).
\end{equation}

The paper is organized as follows. In Sec. 2, we note some well known results about the one- and two-magnon spectrums of the XXZ ferromagnetically polarized chain \eqref{ham}, \eqref{addcond}, and \eqref{ground}. In Sec. 3, within the approach, suggested in Refs.
\cite{14,15} and used in Ref. \cite{10}, we give a finite-$N$ representation for
$\chi(\omega,q,T)$, which, however, is useless, because depends on the unknown total array of finite-$N$ two-magnon wave functions. In Sec. 4, following Ref. \cite{17} and using the {\it exactly known} infinite set of $N=\infty$ two-magnon wave functions we get the explicit integral representation for $\chi(\omega,q,T)$.
In Sec. 5 we find the conditions under which in the easy-axis,
\begin{equation}\label{easyaxis}
|\Delta|>1,
\end{equation}
case the magnon creation contribution to TDSF is separated from the one corresponding to transitions from single magnons to coupled magnon pairs.
Under this separation, we extract $\chi_{\rm magn}(\omega,q,T)$ from $\chi(\omega,q,T)$ and get the mass operator $\Sigma(\omega,q,T)$.
In Sec. 6, we discuss the lineshapes of the resonance contours, presented on the Figs. 1-3, and especially consider the {\it double-low-temperature} (DLT) regime, in which \eqref{lowT} is supplemented by the condition
\begin{equation}\label{2lowT}
\beta E_{\rm w}\gg1.
\end{equation}
Here $E_{\rm w}$ is the magnon band width. Using the Laplace method, we obtain the compact formulas for temperature dependent magnon resonance shift and pseudo-decay rate ("pseudo," because magnons are stable). Within the variety of approaches they have been studied for the variety of models \cite{18,19,20}.
Some formulas of the main text are proved in the Appendix.

\section{One- and two-magnon states}

Introducing the magnon-number operator
\begin{equation}\label{q}
\hat Q\equiv\sum_nQ_n,\qquad Q_n=\frac{1}{2}I-{\bf S}_n^z,
\end{equation}
and using the relations $Q_n|\uparrow_n\rangle=0$ and $Q_n|\downarrow_n\rangle=|\downarrow_n\rangle$,
one decomposes the Fock (physical Hilbert) space corresponding to the ferromagnetic phase \eqref{ground} into the direct sum of $Q$-magnon sectors
${\cal H}={\cal H}_0\oplus{\cal H}_1\oplus\dots$.

The one-dimensional sector ${\cal H}_0$ is generated by $|\emptyset\rangle$, while
the $N$-dimensional one-magnon sector ${\cal H}_1$ is spanned on the Bloch spin waves
\begin{equation}\label{magn1}
|k\rangle={\bf S}^-(-k)|\emptyset\rangle,\qquad{\rm e}^{ikN}=1,
\end{equation}
corresponding to energies
\begin{equation}\label{emagn}
E_{\rm magn}(k)=h+J_z-J_{\bot}\cos{k},
\end{equation}
or, equivalently,
\begin{equation}\label{emagnm}
E_{\rm magn}(k)=h+J_z-|J_{\bot}|\cos{(k-k_{\rm gap})},
\end{equation}
where
\begin{equation}\label{egap}
E_{\rm gap}=E_{\rm magn}(k_{\rm gap}),\qquad J_{\bot}=|J_{\bot}|\cos{k_{\rm gap}}.
\end{equation}
It may be readily proved that the set \eqref{magn1} is complete. Namely,
\begin{equation}\label{compl1N}
\sum_k|k\rangle\langle k|={\tt I}_1=\sum_{n=1}^N{\bf S}^-_n|\emptyset\rangle\langle\emptyset|{\bf S}^+_n,
\end{equation}
where by ${\tt H}_m$ and ${\tt I}_m$ we shall denote the restrictions of the Hamiltonian \eqref{ham} and the unity operator on ${\cal H}_m$.

Following \eqref{emagnm} the magnon band energy width is
\begin{equation}\label{Ew}
E_{\rm w}=2|J_{\bot}|.
\end{equation}

Averaging over ${\cal H}_0\oplus{\cal H}_1$, one gets up to the order $o(\zeta_{\rm g})$
\begin{equation}\label{MN}
2M(T)=1-\frac{2}{N}\sum_k{\rm e}^{-\beta E_{\rm magn}(k)},\qquad{\rm e}^{ikN}=1.
\end{equation}

Let
\begin{equation}\label{magn2}
|k,{\mathfrak n}\rangle=\sum_{n_1<n_2}{\rm e}^{ik(n_1+n_2)/2}\varphi_{n_2-n_1}(k,{\mathfrak n})
{\bf S}^-_{n_1}{\bf S}^-_{n_2}|\emptyset\rangle,
\end{equation}
be the complete orthogonal basis in ${\cal H}_2$. Here $k$ is the crystal momentum (${\rm e}^{ikN}=1$), while the parameter ${\mathfrak n}$ enumerates the set of additional quantum numbers. One has
\begin{eqnarray}\label{Schr2new}
&&{\tt H}_2|k,{\mathfrak n}\rangle=E(k,{\mathfrak n})|k,{\mathfrak n}\rangle,\\\label{solvp1}
&&\sum_{k,\mathfrak n}|k,{\mathfrak n}\rangle\langle k,{\mathfrak n}|={\tt I}_2\nonumber\\
&&=\sum_{1\leq n_1<n_2\leq N}{\bf S}^-_{n_1}{\bf S}^-_{n_2}|\emptyset\rangle\langle\emptyset|
{\bf S}^+_{n_1}{\bf S}^+_{n_2}.
\end{eqnarray}

Solutions of \eqref{Schr2new} and \eqref{solvp1} are explicitly known only at $N=\infty$. Namely, for {\it scattering} and {\it bound} states,
\begin{eqnarray}\label{scatt}
&&\varphi_n^{\rm scatt}(k,\kappa)=\frac{A(k,\kappa){\rm e}^{i\kappa n}-A(k,-\kappa){\rm e}^{-i\kappa n}}{\sqrt{A(k,\kappa)A(k,-\kappa)}},\\\label{bound}
&&\varphi_n^{\rm bound}(k)=\frac{\sqrt{J_z^2-J_{\bot}^2\cos^2{k/2}}}{J_{\bot}\cos{k/2}}\Big(\frac{J_{\bot}}{J_z}\cos{\frac{k}{2}}\Big)^n,\qquad
\end{eqnarray}
the corresponding energies are
\begin{eqnarray}\label{escatt}
&&E_{\rm scatt}(k,\kappa)=2\Big(h+J_z-J_{\bot}\cos{\frac{k}{2}}\cos{\kappa}\Big),\\\label{ebound}
&&E_{\rm bound}(k)=2h+J_z-\frac{J_{\bot}^2}{J_z}\cos^2{\frac{k}{2}}.
\end{eqnarray}
Here in \eqref{scatt}
\begin{equation}\label{a}
A(k,\kappa)\equiv J_{\bot}\cos{\frac{k}{2}}-J_z{\rm e}^{-i\kappa},\qquad \kappa\in(0,\pi).
\end{equation}
Since the bound states should be normalized, \eqref{bound} yields
\begin{equation}\label{boundk}
\Big|\frac{1}{\Delta}\cos{\frac{k}{2}}\Big|<1.
\end{equation}

The completeness condition \eqref{solvp1} takes the form
\begin{widetext}
\begin{equation}\label{compl}
\frac{1}{2\pi}\int_0^{2\pi}dk\Big(\frac{1}{2\pi}\int_0^{\pi}d\kappa|k,\kappa,{\rm scatt}\rangle\langle k,\kappa,{\rm scatt}
+\Theta(J_z^2-J_{\bot}^2\cos^2{k/2})|k,{\rm bound}\rangle\langle k,{\rm bound}|\Big)={\tt I}_2,
\end{equation}
and follows from the relation \cite{15}
\begin{equation}\label{solvp}
\frac{1}{2\pi}\int_0^{\pi}\bar\varphi_m^{\rm scatt}(k,\kappa)\varphi_n^{\rm scatt}(k,\kappa)d\kappa
+\Theta\Big(J_z^2-J_{\bot}^2\cos^2{\frac{k}{2}}\Big)
\bar\varphi_m^{\rm bound}(k)\varphi_n^{\rm bound}(k)=\delta_{m,n}.
\end{equation}
\end{widetext}
Here $\Theta(x)=1$ for $x>0$ and $\Theta(x)=0$ for $x<0$.

According to \eqref{escatt} for each $\kappa$
\begin{equation}\label{scattzone}
E_{\rm scatt}^{\rm down}(k)\leq E_{\rm scatt}(k,\kappa)\leq E_{\rm scatt}^{\rm up}(k),
\end{equation}
where
\begin{eqnarray}\label{scattbounds}
&&E_{\rm scatt}^{\rm down}(k)=2\Big(h+J_z-\Big|J_{\bot}\cos{\frac{k}{2}}\Big|\Big),\nonumber\\
&&E_{\rm scatt}^{\rm up}(k)=2\Big(h+J_z+\Big|J_{\bot}\cos{\frac{k}{2}}\Big|\Big).
\end{eqnarray}

\section{Representation for $\chi(\omega,q,T)$ at $N<\infty$}

According to \eqref{s(q)}
\begin{equation}\label{[sqsq]}
\langle[{\bf S}^+(q),{\bf S}^-(-q)]\rangle=2M(T).
\end{equation}

Accounting for \eqref{[sqsq]}, one readily gets from \eqref{twoeqs}
\begin{eqnarray}\label{pt1}
&&(\omega+i\epsilon)\chi(\omega,q,T)=2M(T)+F(\omega,q,T),\\\label{pt2}
&&(\omega+i\epsilon)F(\omega,q,T)=V(q,T)+W(\omega,q,T),
\end{eqnarray}
where
\begin{widetext}
\begin{eqnarray}\label{vw}
&&F(\omega,q,T)\equiv\langle\langle[{\bf S}^+(q),\hat H],{\bf S}^-(-q)\rangle\rangle
=\langle\langle{\bf S}^+(q),[\hat H,{\bf S}^-(-q)]\rangle\rangle,\nonumber\\
&&V(q,T)\equiv\langle[[{\bf S}^+(q),\hat H],{\bf S}^-(-q)]\rangle
=\langle[{\bf S}^+(q),[\hat H,{\bf S}^-(-q)]]\rangle,\qquad
W(\omega,q,T)\equiv\langle\langle[{\bf S}^+(q),\hat H],[\hat H,{\bf S}^-(-q)]\rangle\rangle.
\end{eqnarray}
Expressing $(\omega+i\epsilon)F(\omega,q,T)\chi(\omega,q,T)$ both from \eqref{pt1} and \eqref{pt2} one gets
\begin{equation}
[V(q,T)+W(\omega,q,T)]\chi(\omega,q,T)=[2M(T)+F(\omega,q,T)]F(\omega,q,T),
\end{equation}
\end{widetext}
or, equivalently,
\begin{equation}\label{=F}
F(\omega,q,T)=\frac{[V(q,T)+W^{(1)}(\omega,q,T)]\chi(\omega,q,T)}{2M(T)},
\end{equation}
where
\begin{equation}\label{w1}
W^{(1)}(\omega,q,T)\equiv W(\omega,q,T)-\frac{F^2(\omega,q,T)}{\chi(\omega,q,T)}.
\end{equation}

The substitution of $F(\omega,q,T)$ from \eqref{=F} into \eqref{pt1} yields the formula
\begin{equation}\label{dys0}
\chi(\omega,q,T)=\cfrac{2M(T)}{\omega-\cfrac{V(q,T)+W^{(1)}(\omega,q,T)}{2M(T)}+i\epsilon},
\end{equation}
very similar to the Dyson equation. However, this analogy is not full, because according to \eqref{w1} $W^{(1)}(\omega,q,T)$ in itself depends on $\chi(\omega,q,T)$. Nevertheless, up to the order $o(\zeta_{\rm g})$ the Dyson equation may be obtained from \eqref{dys0}.

First, let us represent \eqref{w1} in the form
\begin{equation}\label{w1=}
W^{(1)}(\omega,q,T)=\langle\langle[{\bf S}^+(q),\hat H],[\hat H,{\bf S}^-(-q)]\rangle\rangle_1,
\end{equation}
where for two arbitrary operators ${\cal A}$ and ${\cal B}$
\begin{equation}\label{green1}
\langle\langle{\cal A},{\cal B}\rangle\rangle_1\equiv\langle\langle{\cal A},{\cal B}\rangle\rangle-
\frac{\langle\langle{\cal A},{\bf S}^-(-q)\rangle\rangle\langle\langle{\bf S}^+(q),{\cal B}\rangle\rangle}{\langle\langle{\bf S}^+(q),{\bf S}^-(-q)\rangle\rangle}.
\end{equation}
satisfies the {\it irreducibility condition} ($\eta_{\pm}\in{\mathbb C}$),
\begin{equation}\label{irr1}
\langle\langle{\cal A}+\eta_+{\bf S}^+(q),{\cal B}+\eta_-{\bf S}^-(-q)\rangle\rangle_1=\langle\langle{\cal A},{\cal B}\rangle\rangle_1.
\end{equation}
Hence, \eqref{w1=} is equivalent to
\begin{equation}\label{ohho=toto1}
W^{(1)}(\omega,q,T)=\langle\langle X(q),X^{\dagger}(q)\rangle\rangle_1,
\end{equation}
where
\begin{equation}\label{x}
X(q)=[{\bf S}^+(q),\hat H]-E_{\rm magn}(q){\bf S}^+(q).
\end{equation}

At the same time, following \eqref{ground}, \eqref{magn1}, and \eqref{x},
\begin{equation}\label{xvac}
X(q)|\emptyset\rangle=X^{\dagger}(q)|\emptyset\rangle=0.
\end{equation}
Accounting in \eqref{lehm} for \eqref{magn1} and \eqref{xvac}, one gets
\begin{equation}\label{w1=w}
\langle\langle X(q),X^{\dagger}(q)\rangle\rangle_1=\langle\langle X(q),X^{\dagger}(q)\rangle\rangle+o(\zeta_{\rm g}).
\end{equation}

From \eqref{xvac} and \eqref{MN}, one has $\langle\langle X(q),X^{\dagger}(q)\rangle\rangle=O(\zeta_{\rm g})$ and
$2M(T)=1+O(\zeta_{\rm g})$. So, up to the order $o(\zeta_{\rm g})$,
\begin{equation}\label{tw/2s}
\frac{W^{(1)}(\omega,q,T)}{2M(T)}=\langle\langle X(q),X^{\dagger}(q)\rangle\rangle.
\end{equation}
The right-hand side of \eqref{tw/2s} does not depend on $\chi(\omega,q,T)$. Hence, its substitution into \eqref{dys0} results in the low-temperature Dyson
equation whose explicit form may be obtained with the use of {\it only} the one- and two-magnon formfactors. This trick is the {\it keystone} of our approach.

According to \eqref{vw} and \eqref{x},
\begin{equation}\label{V/2M}
\frac{V(q,T)}{2M(T)}=E_{\rm magn}(q)+\varepsilon^{(0)}(q,T),
\end{equation}
where
\begin{equation}\label{e0def}
\varepsilon^{(0)}(q,T)\equiv\frac{\langle[X(q),{\bf S}^-(-q)]\rangle}{2M(T)}
=\frac{\langle[{\bf S}^+(q),X^{\dagger}(q)]\rangle}{2M(T)}.
\end{equation}
Accounting for \eqref{xvac} and \eqref{MN}, one readily gets from \eqref{e0def} up to the order $o(\zeta_{\rm g})$,
\begin{equation}\label{e0red}
\varepsilon^{(0)}(q,T)=\sum_k{\rm e}^{-\beta E_{\rm magn}(k)}\langle k|[X(q),{\bf S}^-(-q)]|k\rangle.
\end{equation}

Following \eqref{lehm} and \eqref{xvac}, up to the order $o(\zeta_{\rm g})$ one has
\begin{equation}\label{XX=}
\langle\langle X(q),X^{\dagger}(q)\rangle\rangle=\chi_1(\omega,q,T),
\end{equation}
where [see \eqref{Schr2new}]
\begin{widetext}
\begin{equation}\label{G1def}
\chi_1(\omega,q,T)\equiv\sum_{k,{\mathfrak n}}\frac{{\rm e}^{-\beta E_{\rm magn}(k-q)}|\langle k,{\mathfrak n}|X^{\dagger}(q)|k-q\rangle|^2}
{\omega+E_{\rm magn}(k-q)-E(k,{\mathfrak n})+i\epsilon},\qquad{\rm e}^{ikN}=1.
\end{equation}
Taking into account that, according to \eqref{magn1} and \eqref{xvac}, $X^{\dagger}(q)|k-q\rangle=[X^{\dagger}(q),{\bf S}^-(q-k)]|\emptyset\rangle$,
one may rewrite \eqref{G1def} as
\begin{equation}\label{G1defmod}
\chi_1(\omega,q,T)=\frac{1}{N}\sum_k{\rm e}^{-\beta E_{\rm magn}(k-q)}
\langle\emptyset|[{\bf S}^+(k-q),X(q)]{\tt G}_N(\omega,q,k)[X^{\dagger}(q),{\bf S}^-(q-k)]|\emptyset\rangle,\qquad{\rm e}^{ikN}=1,
\end{equation}
\end{widetext}
where, following \eqref{Schr2new} and \eqref{solvp1}, the $2^N\times2^M$ matrix
\begin{equation}\label{GNdef}
{\tt G}_N(\omega,q,k)\equiv N\sum_{\mathfrak n}\frac{|k,\mathfrak n\rangle\langle k,\mathfrak n|}{\omega+E_{\rm magn}(k-q)-E(k,{\mathfrak n})+i\epsilon},
\end{equation}
satisfies the equation
\begin{equation}\label{resolvN}
\frac{1}{N}\sum_k(\omega+E_{\rm magn}(k-q)-{\tt H}_2+i\epsilon){\tt G}_N(\omega,q,k)={\tt I}_2.
\end{equation}

The substitutions of \eqref{V/2M} and \eqref{tw/2s} into \eqref{dys0} with the account for \eqref{XX=}, result in
\begin{widetext}
\begin{equation}\label{chi}
\chi(\omega,q,T)=\cfrac{2M(T)}{\omega-E_{\rm magn}(q)-\varepsilon^{(0)}(q,T)-\chi_1(\omega,q,T)+i\epsilon}.
\end{equation}
\end{widetext}

\section{$\chi_1(\omega,q,T)$ at $N=\infty$}

At $N\rightarrow\infty$, under the substitution
\begin{equation}\label{subst}
\frac{1}{N}\sum_{j=1}^N\longrightarrow\frac{1}{2\pi}\int_0^{2\pi}dk
\end{equation}
\eqref{MN} turns into
\begin{equation}\label{M}
2 M(T)=1-\frac{1}{\pi}\int_0^{2\pi}dk{\rm e}^{-\beta E_{\rm magn}(k)}.
\end{equation}
As it is shown in the Appendix A,
\begin{widetext}
\begin{equation}\label{e0}
\varepsilon^{(0)}(q,T)=\frac{1}{\pi}\int_0^{2\pi}dk{\rm e}^{-\beta E_{magn}(k)}[J_{\bot}\cos{q}-J_z
+(J_{\bot}-J_z\cos{q})\cos{k}].
\end{equation}
\end{widetext}

At finite-$N$ the explicit representations for $|k,{\mathfrak n}\rangle$ and $E(k,{\mathfrak n})$ are unknown. Hence, it is not clear how to get
${\tt G}_N(\omega,q,k)$ from \eqref{GNdef}. As the result, \eqref{G1defmod} is useless for direct calculations. Fortunately, the infinite-dimensional operator
\begin{equation}\label{Ginfty}
{\tt G}_{\infty}(\omega,q,k)\equiv\lim_{N\rightarrow\infty}{\tt G}_N(\omega,q,k),
\end{equation}
may be readily obtained within the approach, suggested in Ref. \cite{17}. Indeed, according to \eqref{resolvN}, \eqref{Ginfty}, and \eqref{subst},
\begin{equation}\label{resolvinf}
\int_0^{2\pi}\frac{dk}{2\pi}(\omega+E_{\rm magn}(k-q)-{\tt H}_2+i\epsilon){\tt G}_{\infty}(\omega,q,k)={\tt I}_2.
\end{equation}
Using \eqref{compl}, one readily gets the solution of the \eqref{resolvinf} as
\begin{equation}\label{Gsum}
{\tt G}_{\infty}(\omega,q,k)={\tt G}_{\rm scatt}(\omega,q,k)+{\tt G}_{\rm bound}(\omega,q,k),
\end{equation}
where
\begin{widetext}
\begin{eqnarray}\label{Gexpl}
&&{\tt G}_{\rm scatt}(\omega,q,k)=\frac{1}{2\pi}\int_0^{\pi}d\kappa\frac{|k,\kappa,{\rm scatt}\rangle\langle k,\kappa,{\rm scatt}|}{\omega+E_{\rm magn}(k-q)-E_{\rm scatt}(k,\kappa)+i\epsilon},\nonumber\\
&&{\tt G}_{\rm bound}(\omega,q,k)=\Theta(J_z^2-J_{\bot}^2\cos^2{k/2})
\frac{|k,{\rm bound}\rangle\langle k,{\rm bound}|}{\omega+E_{\rm magn}(k-q)-E_{\rm bound}(k)+i\epsilon}.
\end{eqnarray}
As the result, \eqref{G1defmod} turns into
\begin{equation}\label{chi1sum}
\chi_1(\omega,q,T)=\chi_1^{(\rm scatt)}(\omega,q,T)+\chi_1^{(\rm bound)}(\omega,q,T),
\end{equation}
where
\begin{eqnarray}\label{chi1N}
&&\chi_1^{(\rm scatt)}(\omega,q,T)=\frac{1}{(2\pi)^2}\int_0^{2\pi}dk\int_0^{\pi}d\kappa
\frac{{\rm e}^{-\beta E_{\rm magn}(k-q)}|\langle k,\kappa,{\rm scatt}|[X^{\dagger}(q),{\bf S}^-(q-k)]|\emptyset\rangle|^2}
{\omega-\Phi_{\rm scatt}(q,k,\kappa)+i\epsilon},\nonumber\\
&&\chi_1^{(\rm bound)}(\omega,q,T)=\frac{1}{2\pi}\int_0^{2\pi}dk\frac{{\rm e}^{-\beta E_{\rm magn}(k-q)}|\langle k,{\rm bound}|[X^{\dagger}(q),{\bf S}^-(q-k)]|\emptyset\rangle|^2}
{\omega-\Phi_{\rm bound}(q,k)+i\epsilon}.
\end{eqnarray}
and
\begin{eqnarray}\label{phiscattbound}
&&\Phi_{\rm scatt}(q,k,\kappa)\equiv E_{\rm scatt}(k,\kappa)-E_{\rm magn}(k-q)=h+J_z+J_{\bot}\cos{(k-q)}
-2J_{\bot}\cos{\frac{k}{2}}\cos{\kappa},\nonumber\\
&&\Phi_{\rm bound}(q,k)\equiv E_{\rm bound}(k)-E_{\rm magn}(k-q)=h+J_{\bot}\cos{(k-q)}-\frac{J_{\bot}^2}{J_z}\cos^2{\frac{k}{2}}.
\end{eqnarray}

As it is shown in Appendix, the substitution of \eqref{chi1N} into \eqref{chi1sum} yields
\begin{equation}\label{chi1fact}
\chi_1(\omega,q,T)=\frac{2}{\pi}\int_0^{2\pi}dk\Xi(q,T,k)
\xi(\omega,q,k),
\end{equation}
where
\begin{eqnarray}\label{k11}
&&\Xi(q,T,k)={\rm e}^{-\beta E_{\rm magn}(k-q)}\Big[J_z\cos{\Big(\frac{k}{2}-q\Big)}-J_{\bot}\cos{\frac{k}{2}}\Big]^2,\nonumber\\
&&\xi(\omega,q,k)=\frac{\omega-h+J_z-J_{\bot}\cos{(k-q)}+\Lambda(k,\omega,q)\sqrt{|D(k,\omega,q)|}}{2J_z(\omega-\Phi_{\rm bound}(q,k)+i\epsilon)},\\
\label{D}
&&D(k,\omega,q)=(\omega-\Phi_{\rm down}(q,k))(\omega-\Phi_{\rm up}(q,k)),\\\label{Lambda}
&&\Lambda(k,\omega,q)=\Theta(\Phi_{\rm down}(q,k)-\omega)-\Theta(\omega-\Phi_{\rm up}(q,k))
-i\Theta(\Phi_{\rm up}(q,k)-\omega)\Theta(\omega-\Phi_{\rm down}(q,k)),
\end{eqnarray}
and
\begin{eqnarray}\label{phidownup}
&&\Phi_{\rm down}(q,k)\equiv E^{\rm down}_{\rm scatt}(k)-E_{\rm magn}(k-q)=h+J_z+J_{\bot}\cos{(k-q)}-2\Big|J_{\bot}\cos{\frac{k}{2}}\Big|,\nonumber\\
&&\Phi_{\rm up}(q,k)\equiv E_{\rm scatt}^{\rm up}(k)-E_{\rm magn}(k-q)=h+J_z+J_{\bot}\cos{(k-q)}+2\Big|J_{\bot}\cos{\frac{k}{2}}\Big|.
\end{eqnarray}
\end{widetext}

\section{Extraction of $\chi_{\rm magn}(\omega,q,T)$}

Following \eqref{chi} and \eqref{e0}, condition \eqref{Imneq} reduces to
\begin{equation}\label{Imchi1}
{\rm Im}\chi_1(\omega,q,T)\neq0,
\end{equation}
or, with the account for \eqref{chi1fact}, \eqref{k11}, \eqref{Lambda} and \eqref{phidownup},
\begin{equation}
\omega\in[\omega_{\rm min}(q),\omega_{\rm max}(q)]\cup[\Omega_{\rm min}(q),\Omega_{\rm max}(q)],
\end{equation}
where
\begin{eqnarray}\label{omegaminmaxdef}
&&\omega_{\rm min}(q)=\min_{k\in[0,2\pi]}\Phi_{\rm down}(q,k),\nonumber\\
&&\omega_{\rm max}(q)=\max_{k\in[0,2\pi]}\Phi_{\rm up}(q,k),\\\label{Omegaminmaxdef}
&&\Omega_{\rm min}(q)=\min_{k\in[0,2\pi]}\Phi_{\rm bound}(q,k),\nonumber\\
&&\Omega_{\rm max}(q)=\max_{k\in[0,2\pi]}\Phi_{\rm bound}(q,k).
\end{eqnarray}
A slight generalization of the analysis given in Ref. \cite{10} yields
\begin{eqnarray}\label{omegaminmaxexpl}
&&\omega_{\rm min}(q)=h+J_z-3|J_{\bot}|\cos\frac{|q|-k_{\rm max}}{3},\qquad|q|\leq\pi,\nonumber\\
&&\omega_{\rm max}(q)=h+J_z+3|J_{\bot}|\cos\frac{|q|-k_{\rm gap}}{3},
\end{eqnarray}
where $k_{\rm max}\equiv\pi-k_{\rm gap}$ is the crystal momentum corresponding to the highest magnon energy.
Using \eqref{emagnm} and \eqref{Ew} and the identity $\cos{3x}=4\cos^3{x}-3\cos{x}$, one may reduce \eqref{omegaminmaxexpl} to the more
tractable form
\begin{eqnarray}\label{omegaminmaxexplm}
&&\omega_{\rm min}(q)=E_{\rm magn}(q)-2E_{\rm w}\cos^3\frac{|q|-k_{\rm max}}{3},\quad|q|\leq\pi,\nonumber\\
&&\omega_{\rm max}(q)=E_{\rm magn}(q)+2E_{\rm w}\cos^3\frac{|q|-k_{\rm gap}}{3}.
\end{eqnarray}
According to \eqref{omegaminmaxexplm}, the condition \eqref{emagn[]} is satisfied automatically, so
the interval $[\omega_{\rm min}(q),\omega_{\rm max}(q)]$ really corresponds to the one-magnon peak.
At the same time, following \eqref{phiscattbound}, the interval $[\Omega_{\rm min}(q),\Omega_{\rm max}(q)]$ corresponds to transitions from scattering magnons to coupled magnon pairs. In order to avoid an account of this rater complex process, we shall look only for the case
\begin{equation}\label{cap}
[\omega_{\rm min}(q),\omega_{\rm max}(q)]\cap[\Omega_{\rm min}(q),\Omega_{\rm max}(q)]=\emptyset.
\end{equation}

Under the easy-axis condition \eqref{easyaxis}, when \eqref{boundk} is satisfied automatically, the explicit
expressions for \eqref{Omegaminmaxdef} may be readily obtained.
Following \eqref{phiscattbound}, the equation $\partial\Phi_{\rm bound}(q,k)/\partial k=0$ has two solutions $k_+$ and $k_-$, for which
\begin{eqnarray}\label{kpm}
\sin{k_{\pm}}=\pm\frac{2J_z\sin{q}}{\sqrt{(2J_z-J_{\bot})^2+8J_zJ_{\bot}\sin^2{q/2}}},\nonumber\\
\cos{k_{\pm}}=\pm\frac{2J_z\cos{q}-J_{\bot}}{\sqrt{(2J_z-J_{\bot})^2+8J_zJ_{\bot}\sin^2{q/2}}}.
\end{eqnarray}
The substitution of \eqref{kpm} into \eqref{phiscattbound} yields
\begin{widetext}
\begin{equation}\label{phiboundkpm}
\Phi_{\rm bound}(q,k_{\pm})=h-\frac{J_{\bot}^2}{2J_z}
\pm\frac{J_{\bot}}{2J_z}\sqrt{(2J_z-J_{\bot})^2+8J_zJ_{\bot}\sin^2{q/2}}.
\end{equation}
Using \eqref{Delta} and \eqref{Ew} and the notations $E_{\rm mid}\equiv h+J_z$ and $E_{\rm bound}^{\rm mid}\equiv2h+J_z-J_{\bot}^2/(2J_z)$ for the middles of magnon and coupled pairs zones, one readily gets from \eqref{phiboundkpm}
\begin{eqnarray}\label{Omegaminmaxexpl}
&&\Omega_{\rm min}(q)=E_{\rm bound}^{\rm mid}-E_{\rm mid}-\frac{E_{\rm w}}{2}\sqrt{1+\frac{1}{4\Delta^2}-\frac{\cos{q}}{\Delta}},\nonumber\\
&&\Omega_{\rm max}(q)=E_{\rm bound}^{\rm mid}-E_{\rm mid}+\frac{E_{\rm w}}{2}\sqrt{1+\frac{1}{4\Delta^2}-\frac{\cos{q}}{\Delta}}.
\end{eqnarray}
\end{widetext}

As it readily follows from \eqref{omegaminmaxexpl} and \eqref{Omegaminmaxexpl}
\begin{eqnarray}\label{limw}
&&\lim_{J_{\bot}\rightarrow0}[\omega_{\rm min}(q),\omega_{\rm max}(q)]=\{h+J_z\},\nonumber\\
&&\lim_{J_{\bot}\rightarrow0}[\Omega_{\rm min}(q),\Omega_{\rm max}(q)]=\{h\}.
\end{eqnarray}
Hence, \eqref{cap} always is satisfied in the Ising-like regime $|\Delta|\gg1$. Following \eqref{limw} we conjecture that the condition \eqref{cap} may be rewritten as
\begin{subequations}\label{capexpand}
\begin{eqnarray}
&&J_z>0:\qquad\omega_{\rm min}(q)-\Omega_{\rm max}(q)>0,\\
&&J_z<0:\qquad\Omega_{\rm min}(q)-\omega_{\rm max}(q)>0.
\end{eqnarray}
\end{subequations}
In the following the system \eqref{easyaxis} and \eqref{capexpand} always should be implied.

The substitution of \eqref{k11} into \eqref{chi1fact}, and accounting for \eqref{dys} and \eqref{self}, yields
\begin{equation}\label{varepsilonsum}
\varepsilon(\omega,q,T)=\varepsilon^{(0)}(q,T)+\sum_{j=1}^2\varepsilon^{(j)}(\omega,q,T),
\end{equation}
and
\begin{widetext}
\begin{eqnarray}\label{e1}
&&\varepsilon^{(1)}(\omega,q,T)=\frac{1}{\pi J_z}\int_0^{2\pi}dk\Xi(q,T,k)
\frac{\omega-h+J_z-J_{\bot}\cos{(k-q)}}
{\omega-\Phi_{\rm bound}(q,k)},\\\label{e2}
&&\varepsilon^{(2)}(\omega,q,T)=\frac{1}{\pi J_z}\int_0^{2\pi}dk\Xi(q,T,k)
\frac{\sqrt{D(\omega,q,k)}}{\omega-\Phi_{\rm bound}(q,k)}\Big(\Theta(\Phi_{\rm down}(q,k)-\omega)-\Theta(\omega-\Phi_{\rm up}(q,k))\Big),\\
\label{Gammag}
&&\Gamma(\omega,q,T)=\frac{1}{\pi J_z}\int_0^{2\pi}dk\Xi(q,T,k)
\frac{\sqrt{-D(\omega,q,k)}}{\omega-\Phi_{\rm bound}(q,k)}
\Theta(\Phi_{\rm up}(q,k)-\omega)\Theta(\omega-\Phi_{\rm down}(q,k)).
\end{eqnarray}

Formula \eqref{e2} may be represented in a more tractable and expanded form. Following \eqref{emagnm} and \eqref{phidownup}
\begin{subequations}\label{capred}
\begin{eqnarray}
&&E_{\rm magn}(q)-\Phi_{\rm down}(q,k)=2|J_{\bot}|\Big[\Big|\cos{\frac{k}{2}}\Big|-\cos{\frac{k}{2}}\cos{\Big(\frac{k}{2}-q-k_{\rm gap}\Big)}\Big]\geq0,\\
&&\Phi_{\rm up}(q,k)-E_{\rm magn}(q)=2|J_{\bot}|\Big[\Big|\cos{\frac{k}{2}}\Big|+\cos{\frac{k}{2}}\cos{\Big(\frac{k}{2}-q-k_{\rm gap}\Big)}\Big]\geq0.
\end{eqnarray}
\end{subequations}
Hence,
\begin{equation}\label{thetaemagn}
\Theta(\omega-\Phi_{\rm up}(q,k))|_{\omega<E_{\rm magn}(q)}=0,\qquad\Theta(\Phi_{\rm down}(q,k)-\omega)|_{\omega>E_{\rm magn}(q)}=0,
\end{equation}
and, as the result, \eqref{e2} splits into two separated formulas
\begin{subequations}\label{e2red}
\begin{eqnarray}
&&\varepsilon^{(2)}(\omega,q,T)=\frac{1}{\pi J_z}\int_0^{2\pi}dk\Xi(q,T,k)
\frac{\Theta(\Phi_{\rm down}(q,k)-\omega)\sqrt{-D(\omega,q,k)}}{\omega-\Phi_{\rm bound}(q,k)},\qquad\omega\in[\omega_{\rm min}(q),E_{\rm magn}(q)],\\
&&\varepsilon^{(2)}(\omega,q,T)=-\frac{1}{\pi J_z}\int_0^{2\pi}dk\Xi(q,T,k)
\frac{\Theta(\omega-\Phi_{\rm up}(q,k))\sqrt{-D(\omega,q,k)}}{\omega-\Phi_{\rm bound}(q,k)},\qquad
\omega\in[E_{\rm magn}(q),\omega_{\rm max}(q)].
\end{eqnarray}
\end{subequations}

Finally, the combination of \eqref{tdsfmagn}, \eqref{dys}, and \eqref{self} yields [up to the order $o(\zeta_{\rm g})$]
\begin{equation}\label{result}
S_{\rm magn}(\omega,q,T)=\frac{1}{\pi(1-{\rm e}^{-\beta\omega})}\cdot
\frac{2M(T)\Gamma(\omega,q,T)}{(\omega-E_{\rm magn}(q)-\varepsilon(\omega,q,T))^2+\Gamma^2(\omega,q,T)}.
\end{equation}
\end{widetext}

\section{Asymmetry and broadening of the resonance contour lineshapes}

For the ferromagnetic chain with $h=3$, $J_z=1$, and $J_{\bot}=0.2$ ($E_{\rm gap}=3.8$, $E_{\rm w}=0.4$, $\Delta=5$) some resonance contours at $q=0,\,\pi/2,\,\pi$  are presented in the Figs. 1-3, where it is implied that $k_B=1$ ($\beta=1/T$). In these figures
\begin{equation}
E_{\rm magn}(q,T)\equiv E_{\rm magn}(q)+\varepsilon_{\rm magn}(q,T),
\end{equation}
where the magnon resonance shift $\varepsilon_{\rm magn}(q,T)$ [the difference between the lineshape maximum and $E_{\rm magn}(q)$] is evaluated according to the approximative formula \eqref{e=e0+e1} supplemented by \eqref{e0} and \eqref{e1}.
The presented plots have rather custom lineshapes which are broadened and asymmetric. The broadening increase with temperature.
The asymmetry may be alternatively characterized by different left and right spreadings around the point $\omega=E_{\rm magn}(q)$ or by the
resonance shift $\varepsilon_{\rm magn}(q,T)$ mentioned above.

The spreading asymmetry is well estimated by the universal ($T$-independent) function
\begin{equation}\label{as}
{\tt as}(q)\equiv\frac{\omega_{\rm max}(q)-E_{\rm magn}(q)}{E_{\rm magn}(q)-\omega_{\rm min}(q)}=
\frac{\cos^3{\displaystyle\frac{|q|-k_{\rm gap}}{3}}}{\cos^3{\displaystyle\frac{|q|-k_{\rm max}}{3}}},
\end{equation}
whose plot is presented in the Fig. 4. As it follows from \eqref{as}, ${\tt as}(k_{\rm gap})=8$ (the contour spreads to the right),
${\tt as}(k_{\rm max})=0.125$ (the contour spreads to the left), and ${\tt as}(\pi/2)=1$ (the contour equally spreads to the right and to the left).

At low temperatures the parameter $\varepsilon_{\rm magn}(q,T)$ with a good accuracy may be obtained from $\omega_c$, the complex pole of $S_{\rm magn}(\omega,q,T)$. Namely
\begin{equation}\label{wc}
\omega_c\equiv E_{\rm magn}(q)+\varepsilon_{\rm magn}(q,T)-i\Gamma_{\rm magn}(q,T),
\end{equation}
where, following \eqref{result},
\begin{equation}\label{pole}
\omega_c-E_{\rm magn}(q)-\varepsilon(\omega_c,q,T)+i\Gamma(\omega_c,q,T)=0.
\end{equation}

Although the parameter $\Gamma_{\rm magn}(q,T)$ corresponds to the lineshape broadening, its interpretation as the magnon decay rate is incorrect, because within the model \eqref{ham} magnons are stable. At $T>0$ the inequality $\Gamma_{\rm magn}(q,T)>0$ follows from the fact that (contrary to ${\bf S}^-(-q)|\emptyset\rangle$) vector  ${\bf S}^-(-q)|k\rangle$ is not an eigenstate of \eqref{ham} but an infinite linear combination of eigenstates with crystal momentums equal to $k+q$.

Up to the order $o(\zeta_{\rm g})$ one has from \eqref{wc} and \eqref{pole}
\begin{eqnarray}\label{EGamma}
&&\varepsilon_{\rm magn}(q,T)=\varepsilon(E_{\rm magn}(q),q,T),\nonumber\\
&&\Gamma_{\rm magn}(q,T)=\Gamma(E_{\rm magn}(q),q,T).
\end{eqnarray}
According to \eqref{EGamma} the exact integral representations for $\varepsilon_{\rm magn}(q,T)$ and $\Gamma_{\rm magn}(q,T)$ may be obtained by the substitution $\omega=E_{\rm magn}(q)$ into \eqref{e1}, \eqref{e2}, and \eqref{Gammag}.
Following \eqref{D}, \eqref{capred}, and \eqref{e2},
\begin{eqnarray}\label{De2}
&&-D(k,E_{\rm magn}(q),q)=4J_{\bot}^2\cos^2{\frac{k}{2}}\sin^2{\Big(\frac{k}{2}-q\Big)},\nonumber\\
&&\varepsilon^{(2)}(E_{\rm magn}(q),q,T)=0.
\end{eqnarray}
Accounting for \eqref{De2}, one readily gets from \eqref{EGamma},
\begin{equation}\label{e=e0+e1}
\varepsilon_{\rm magn}(q,T)=\varepsilon^{(0)}(q,T)+\varepsilon^{(1)}_{\rm magn}(q,T),
\end{equation}
where $\varepsilon^{(0)}(q,T)$ is given by \eqref{e0} and
\begin{widetext}
\begin{eqnarray}\label{e1shift}
&&\varepsilon_{\rm magn}^{(1)}(q,T)=\frac{2}{\pi}\int_0^{2\pi}dk\Xi(q,T,k)
\frac{J_z-J_{\bot}\cos{k/2}\cos{\alpha}}
{[J_z-J_{\bot}\cos{k/2}\cos{\alpha}]^2+J_{\bot}^2\cos^2{k/2}\sin^2{\alpha}},\\\label{G}
&&\Gamma_{\rm magn}(q,T)=\frac{2}{\pi}\int_0^{2\pi}dk\Xi(q,T,k)
\frac{|J_{\bot}\cos{k/2}\sin{\alpha}|}
{[J_z-J_{\bot}\cos{k/2}\cos{\alpha}]^2+J_{\bot}^2\cos^2{k/2}\sin^2{\alpha}},\qquad\alpha\equiv\frac{k}{2}-q.
\end{eqnarray}

It is convenient to express $\varepsilon^{(0)}(q,T)$, $\varepsilon_{\rm magn}^{(1)}(q,T)$, and $\Gamma_{\rm magn}(q,T)$ in terms of the experimentally observable parameters $E_{\rm gap}$, $E_{\rm w}$, and $\Delta$. Using \eqref{Delta}, \eqref{emagnm}, \eqref{egap}, and \eqref{Ew} and the identity $\cos{(q-k_{\rm gap})}=\cos{k_{\rm gap}}\cos{q}$, one readily rewrites \eqref{e0}, \eqref{e1shift}, and \eqref{G} as
\begin{eqnarray}\label{e0e1}
&&\varepsilon^{(0)}(q,T)=\frac{E_{\rm w}\zeta_{\rm g}\cos{k_{\rm gap}}}{2\pi}\int_0^{2\pi}dk{\rm e}^{-\beta E_{\rm w}\sin^2{[(k-k_{\rm gap})/2]}}
[\cos{q}-\Delta
+(1-\Delta\cos{q})\cos{k}],\nonumber\\
&&\varepsilon_{\rm magn}^{(1)}(q,T)=\frac{E_{\rm w}\zeta_{\rm g}\cos{k_{\rm gap}}}{\pi}\int_0^{2\pi}dk
\frac{{\rm e}^{-\beta E_{\rm w}\sin^2{[(k-q-k_{\rm gap})/2]}}(\Delta\cos{\alpha}-\cos{k/2})^2(\Delta-\cos{k/2}\cos{\alpha})}
{(\Delta-\cos{k/2}\cos{\alpha})^2+\cos^2{k/2}\sin^2{\alpha}},\\\label{Gint}
&&\Gamma_{\rm magn}(q,T)=\frac{E_{\rm w}\zeta_{\rm g}}{\pi}\int_0^{2\pi}dk
\frac{{\rm e}^{-\beta E_{\rm w}\sin^2{[(k-q-k_{\rm gap})/2]}}(\Delta\cos{\alpha}-\cos{k/2})^2|\cos{k/2}\sin{\alpha}|}
{(\Delta-\cos{k/2}\cos{\alpha})^2+\cos^2{k/2}\sin^2{\alpha}}.
\end{eqnarray}

Taking the integrals in \eqref{e0e1} within the Laplace method and using the identities
\begin{eqnarray}
&&1-\Delta\cos{k_{\rm gap}}+(\cos{k_{\rm gap}}-\Delta)\cos{q}=(\cos{k_{\rm gap}}-\Delta)(\cos{q}+\cos{k_{\rm gap}}),\nonumber\\
&&\cos^2\frac{q+k_{\rm gap}}{2}\sin^2\frac{q-k_{\rm gap}}{2}=\frac{1}{4}\sin^2q,\qquad\cos\frac{q+k_{\rm gap}}{2}\cos\frac{q-k_{\rm gap}}{2}
=\frac{\cos{q}+\cos{k_{\rm gap}}}{2},\nonumber\\
&&\Big(\Delta\cos{\frac{k_{\rm gap}-q}{2}}-\cos{\frac{k_{\rm gap}+q}{2}}\Big)^2=\frac{(\Delta-\cos{k_{\rm gap}})^2(1+\cos{k_{\rm gap}}\cos{q})}{2},
\end{eqnarray}
one readily gets in the DLT regime \eqref{lowT} and \eqref{2lowT}
\begin{eqnarray}\label{e0e1DLT}
&&\varepsilon_{\rm magn}^{(0){\rm DLT}}(q,T)=\zeta_{\rm DLT}E_{\rm w}(\cos{k_{\rm gap}}-\Delta)(\cos{q}+\cos{k_{\rm gap}}),\nonumber\\
&&\varepsilon_{\rm magn}^{(1){\rm DLT}}(q,T)=
\frac{2\zeta_{\rm DLT}E_{\rm w}(\cos{k_{\rm gap}}-\Delta)^2(\cos{q}+\cos{k_{\rm gap}})(2\Delta-\cos{q}-\cos{k_{\rm gap}})}{(2\Delta-\cos{q}-\cos{k_{\rm gap}})^2+\sin^2{q}},\\\label{eDLT}
&&\varepsilon_{\rm magn}^{{\rm DLT}}(q,T)=\frac{2\Delta\zeta_{\rm DLT}E_{\rm w}(\cos{k_{\rm gap}}-\Delta)(\cos{q}+\cos{k_{\rm gap}})\sin^2{q}}{(2\Delta-\cos{q}-\cos{k_{\rm gap}})^2+\sin^2{q}},
\end{eqnarray}
\end{widetext}
where
\begin{equation}
\zeta_{\rm DLT}\equiv\frac{\zeta_{\rm g}}{\sqrt{\pi\beta E_{\rm w}}}.
\end{equation}

At $q=k_{\rm gap}$ and $q=k_{\rm max}$ ($\sin{q}=0$) the right-hand side of \eqref{eDLT} reduces. Hence, in order to get a nontrivial result, one have
to use the formula $\int_{-\infty}^{\infty}x^2{\rm e}^{-x^2}dx=\sqrt{\pi}/2$ (instead of $\int_{-\infty}^{\infty}{\rm e}^{-x^2}dx=\sqrt{\pi}$).
As shown in the Appendix, this approach yields
\begin{eqnarray}\label{e(kgap)}
&&\varepsilon_{\rm magn}^{\rm DLT}(k_{\rm gap},T)=\frac{\zeta_{\rm DLT}\Delta k_BT}{(\cos{k_{\rm gap}}-\Delta)},\\\label{e(kmax)}
&&\varepsilon_{\rm magn}^{\rm DLT}(k_{\rm max},T)=\frac{\zeta_{\rm DLT}(\Delta-\cos{k_{\rm max}})k_BT}{\Delta}.
\end{eqnarray}

At vicinities of the points $q=k_{\rm gap}$ and $q=k_{\rm max}$ one has to account both for \eqref{eDLT} and \eqref{e(kgap)} or \eqref{e(kmax)}.
The joint utilization of these formulas gives
\begin{widetext}
\begin{eqnarray}\label{kgapmax}
&&\varepsilon_{\rm magn}^{\rm DLT}(q,T)=\frac{\zeta_{\rm g}\Delta(2\cos{k_{\rm gap}}E_{\rm w}(q-k_{\rm gap})^2+k_BT)}
{2(\cos{k_{\rm gap}}-\Delta)},\qquad q\approx k_{\rm gap}\nonumber\\
&&\varepsilon_{\rm magn}^{\rm DLT}(q,T)=\frac{\zeta_{\rm g}[E_{\rm w}(1+\Delta\cos{k_{\rm max}})(q-k_{\rm max})^4
+2(\Delta-\cos{k_{\rm max}})k_BT]}
{4\Delta},\qquad q\approx k_{\rm max}.
\end{eqnarray}
\end{widetext}

Following \eqref{easyaxis}, \eqref{eDLT} and \eqref{kgapmax} at $T\rightarrow0$ in the DLT regime outside from the points $q=k_{\rm gap}$ and $q=k_{\rm max}$ the sign of $\varepsilon_{\rm magn}^{\rm DLT}(q,T)$ is equal to the sign of
$\cos{k_{\rm max}}=-\cos{k_{\rm gap}}=-{\rm sign}(J_{\bot})$. At the same time, in the contours presented in the Fig. 2 all the shifts are positive, though $q=\pi/2$ and $k_{\rm gap}=0$. One may suppose, that this is a consequence of insufficient low temperatures. Namely, for $h=3$, $J_z=1$ and $J_{\bot}=0.2$ one has $E_{\rm gap}=3.8$ and $E_{\rm w}=0.4$. Hence, ${\rm e}^{-\beta E_{\rm gap}}=0.2$ even for $T=2.5$. So the approximation governed by \eqref{lowT} should be correct. At the same time, $\beta E_{\rm w}=3.25$ even for $T=1.3$, so that fulfillment of the condition \eqref{2lowT} is
rather doubtful. In order to clarify the question we illustrate the temperature dependence of $\varepsilon_{\rm magn}(\pi/2,T)$ in the Fig. 5. As we can see up to $T=0.02$ ($\beta E_{\rm w}=20$) the plots of $\varepsilon_{\rm magn}(\pi/2,T)$ and $\varepsilon_{\rm magn}^{\rm DLT}(\pi/2,T)$ are practically indistinguishable while up to $T=0.43$ $(\beta E_{\rm w}=10.7)$ both of them are negative. The point $T=1.3$ lies far beyond the DLT regime.

Using the plots presented in the Figs. 1-3, one may also estimate the low-temperature approximation used in \eqref{EGamma}. As it may be readily seen, the difference between $E_{\rm magn}(q,T)$ and the lineshape maximum is rather negligible only for $T=1.3$.

\section{Summary and discussion}

In the present paper, following the line of research suggested in Ref. \cite{10}, we have evaluated the magnon creation contribution to the transverse dynamical structure factor (TDSF) for a ferromagnetically polarized, {\it gapped}, easy-axis
[the latter condition \eqref{easyaxis}, supplemented by \eqref{cap} or \eqref{capexpand} enables the separation between magnon- and coupled-pair- creation contributions] XXZ chain \eqref{ham} up to the order $o({\rm e}^{-\beta E_{\rm gap}})$. The final result \eqref{result} was obtained
according to the well known correspondence between TDSF and the corresponding transverse dynamical magnetic susceptibility.
The latter was presented in the form of the Dyson equation \eqref{dys} with the use of the Plakida-Tserkovnikov approach \cite{13}, \cite{15} supplemented by the special low-temperature reduction suggested by the author. All the calculations were implemented
with utilization of {\it only one- and two-magnon spectrums}.

Treating the mass operator \eqref{self}, we have got the rather tractable integral representations \eqref{e=e0+e1}, \eqref{e0e1}, and \eqref{Gint} for the temperature-dependent magnon resonance shift $\varepsilon_{\rm magn}(q,T)$ and "decay rate" $\Gamma_{\rm magn}(q,T)$. Contrary to the heuristical approaches, used in Refs. \cite{3,21}, where TDSF was evaluated by ad hoc substitution of {\it finite} $\epsilon$ into the spectral formula \eqref{lehm},
the suggested one is well grounded and produces estimations with the {\it controllable} order $o({\rm e}^{-\beta E_{\rm gap}})$.
In the special double low temperature (DLT) regime, when temperature is small both up to the energy gap and to the magnon band width, some of the obtained integral representations has been evaluated within the Laplace method. The correctness of this approximation was estimated by comparison with the reference result.

It was demonstrated that inside the DLT regime $\varepsilon_{\rm magn}(q,T)$ has the sign opposite to the sign of $\cos{k_{\rm gap}}$. However,
beyond the DLT as a function of $T$ it may have a rather nontrivial behavior.

At present time a number of magnetic compounds related to the model \eqref{ham}, \eqref{ground}, and \eqref{easyaxis} are known. Among them are both ferromagnets \cite{22,23} and magnetically polarized antiferromagnets \cite{24}. To the authors knowledge, the experimental TDSF data for them
was not published.
Moreover, in the author's opinion, it is premature now to use the presented results for interpretation of experimental data \cite{11},
because a discrepancy between theory and experiment may occur as well as from the suggested approximations (utilization of only one- and
two-magnon spectrums), as well as from an incorrectness of the reference model.
That is why, it seems for the author, that the obtained results should be initially confirmed numerically.

Since the late 2000s several effective numerical approaches for the high accuracy evaluation of temperature-dependent dynamical correlations in spin chains were developed. Among them are an exact diagonalization in finite chains \cite{7,9}, quantum Monte Carlo methods \cite{25}, and the Density Matrix Renormalization Group approach \cite{26,27,28}.
Within these methods several systems, more complex than \eqref{ham} and \eqref{ground}, were studied. Unfortunately, just the TDSF for the model \eqref{ham}, \eqref{ground}, and \eqref{easyaxis} has not yet been considered.
Since the suggested approach gives a rather complete information about the magnon resonance contour lineshape, it seems reasonable to select the most physically interesting quantities and after a combined analytical and numerical study formulate recommendations for experimentalists.

The suggested approach may be applied to other models with known
one- and two-magnon spectrums \cite{29,30}.

The author is very grateful to S. B. Rutkevich for the helpful discussion.

\appendix
\renewcommand{\theequation}{\thesection.\arabic{equation}}
\section{}
\setcounter{equation}{0}
\begin{itemize}
\item

According to \eqref{s(q)}, \eqref{ham}, and \eqref{x},
\begin{widetext}
\begin{eqnarray}\label{xexpl}
&&X(q)=\frac{1}{\sqrt{N}}\sum_{n=1}^N{\rm e}^{-iqn}{\bf S}^+_n\Big[(J_{\bot}{\rm e}^{iq}-J_z)Q_{n-1}
+(J_{\bot}{\rm e}^{-iq}-J_z)Q_{n+1}\Big],\\\label{[XS(q)]}
&&[X(q),{\bf S}^-(-q)]=\frac{1}{N}\sum_{n=1}^N\Big[2{\bf S}^z_n\Big((J_{\bot}{\rm e}^{iq}-J_z)Q_{n-1}
+(J_{\bot}{\rm e}^{-iq}-J_z)Q_{n+1}\Big)\nonumber\\
&&+{\bf S}^+_n\Big((J_{\bot}-J_z{\rm e}^{-iq}){\bf S}^-_{n-1}
+(J_{\bot}-J_z{\rm e}^{iq}){\bf S}^-_{n+1}\Big)\Big],\\\label{[XS]}
&&[X^{\dagger}(q),{\bf S}^-(q-k)]
=\frac{2}{N}\sum_{n=1}^N{\rm e}^{ik(n+1/2)}\Big[J_z\cos{\Big(\frac{k}{2}-q\Big)}
-J_{\bot}\cos{\frac{k}{2}}\Big]{\bf S}_n^-{\bf S}_{n+1}^-,
\end{eqnarray}
Accounting for \eqref{[XS(q)]}, \eqref{magn1}, \eqref{ground}, and \eqref{magn2} one readily gets from \eqref{e0red}
and \eqref{[XS]}
\begin{eqnarray}\label{e0N}
&&\varepsilon^{(0)}(q,T)=\frac{2}{N}\sum_k{\rm e}^{-\beta E_{\rm magn}(k)}[J_{\bot}(\cos{q}+\cos{k})
-J_z(\cos{(k-q)+1)}],\\\label{matrixelement}
&&\langle k,{\mathfrak n}|[X^{\dagger}(q),{\bf S}^-(q-k)]|\emptyset\rangle=2{\rm e}^{ik/2}\Big(J_z\cos{\Big(\frac{k}{2}-q\Big)}-J_{\bot}\cos{\frac{k}{2}}\Big)
\bar\varphi_1(k,{\mathfrak n}).
\end{eqnarray}

Under the identity $\int_0^{2\pi}{\rm e}^{-\beta E_{magn}(k)}\sin{k}dk=0$, from \eqref{e0N} follows \eqref{e0}.
According to \eqref{matrixelement}, \eqref{chi1N} turns into \eqref{chi1fact},
where
\begin{equation}\label{xisum}
\xi(\omega,q,k)=\xi_{\rm scatt}(\omega,q,k)+\xi_{\rm bound}(\omega,q,k),
\end{equation}
and
\begin{equation}\label{xiscattbound}
\xi_{\rm scatt}(\omega,q,k)=\frac{1}{2\pi}\int_0^{\pi}
\frac{|\varphi_1^{\rm scatt}(k,\kappa)|^2d\kappa}
{\omega-\Phi_{\rm scatt}(q,k,\kappa)+i\epsilon},\qquad
\xi_{\rm bound}(\omega,q,k)=\frac{\Theta(J_z^2-J_{\bot}^2\cos^2{k/2})|\varphi_1^{\rm bound}(k,\kappa)|^2}
{\omega-\Phi_{\rm bound}(q,k,\kappa)+i\epsilon}.
\end{equation}

\item

Using invariance of $\Phi_{\rm scatt}(q,k,\kappa)$ under the substitution $\kappa\rightarrow-\kappa$ and an explicit representation
\begin{equation}\label{xiexpl}
\bar\varphi_1^{\rm scatt}(k,\kappa)\varphi_1^{\rm scatt}(k,\kappa)=\frac{A(k,-\kappa)-A(k,\kappa){\rm e}^{2i\kappa}}{A(k,-\kappa)}
+\frac{A(k,\kappa)-A(k,-\kappa){\rm e}^{-2i\kappa}}{A(k,\kappa)},
\end{equation}
which directly follows from \eqref{scatt} and \eqref{a}, one may extend the integral in \eqref{xiscattbound} into the integral over $[0,2\pi]$,
\begin{equation}\label{xiscatt}
\xi_{\rm scatt}(\omega,q,k)
=\frac{1}{2\pi}\int_0^{2\pi}d\kappa
\frac{A(k,-\kappa)-A(k,\kappa){\rm e}^{2i\kappa}}{A(k,-\kappa)(\omega-\Phi_{\rm scatt}(q,k,\kappa)+i\epsilon)}.
\end{equation}

Introducing the following auxiliary variables,
\begin{equation}\label{auxilir}
a\equiv J_{\bot}\cos{\frac{k}{2}},\qquad z_0\equiv\frac{a}{J_z},\qquad z\equiv{\rm e}^{i\kappa},\qquad
b\equiv \omega-h-J_z-J_{\bot}\cos{(k-q)},
\end{equation}
one readily gets from \eqref{phidownup}, \eqref{phiscattbound}, and \eqref{a},
\begin{eqnarray}\label{omega1}
&&\omega-\Phi_{\rm down}(q,k)=b+2|a|,\qquad
\omega-\Phi_{\rm up}(q,k)=b-2|a|,\\\label{omega2}
&&\omega-\Phi_{\rm scatt}(q,k,\kappa)=a(z+z^{-1})+b,\qquad\frac{A(k,-\kappa)-A(k,\kappa){\rm e}^{2i\kappa}}{A(k,-\kappa)}=\frac{z_0(z^2-1)}{z-z_0},\\\label{omega3}
&&\omega-\Phi_{\rm bound}(q,k)=a(z_0+z^{-1}_0)+b.
\end{eqnarray}

The substitution of \eqref{omega2} into \eqref{xiscatt} yields
\begin{equation}\label{xiscatt2}
\xi_{\rm scatt}(\omega,q,k)=
\frac{z_0}{2\pi i}\oint_{|z|=1}\frac{(z^2-1)dz}{[a(z^2+1)+(b+i\epsilon)z](z-z_0)}
=\frac{z_0(z_{\rm in}^2-1)}{a(z_{\rm in}-z_{\rm out})(z_{\rm in}-z_0)}
+\frac{\Theta(1-z_0^2)(z_0^2-1)}{a(z_0+z_0^{-1})+b+i\epsilon},
\end{equation}
\end{widetext}
where
\begin{equation}\label{zinout}
z_{\rm in}z_{\rm out}=1,\qquad |z_{\rm in}|<1,\qquad|z_{\rm out}|>1,
\end{equation}
are the two roots of the equation
\begin{equation}\label{eqonz}
a(z^2+1)+(b+i\epsilon)z=0.
\end{equation}
At the same time, following \eqref{xiscattbound}, \eqref{bound}, and \eqref{omega3},
\begin{equation}\label{xibound}
\xi_{\rm bound}(\omega,q,k)=\frac{\Theta(1-z_0^2)(1-z_0^2)}{a(z_0+z^{-1}_0)+b+i\epsilon},
\end{equation}
The substitutions of \eqref{xiscatt2} and \eqref{xibound} into \eqref{xisum} yield
\begin{equation}\label{xi0}
\xi(\omega,q,k)=\frac{z_0(z_{\rm in}^2-1)}{a(z_{\rm in}-z_{\rm out})(z_{\rm in}-z_0)}.
\end{equation}
Accounting now for \eqref{omega3} and the identities
\begin{eqnarray}
&&\frac{z_{\rm in}^2-1}{z_{\rm in}-z_{\rm out}}=\frac{z_{\rm in}(z_{\rm in}-z_{\rm out})}{z_{\rm in}-z_{\rm out}}=z_{\rm in},\nonumber\\
&&\frac{z_0}{z_{\rm in}-z_0}=\frac{z_0(z_{\rm out}-z_0)}{(z_{\rm in}-z_0)(z_{\rm out}-z_0)}\nonumber\\
&&=\frac{z_{\rm out}-z_0}{a(z_0+z_0^{-1})+b+i\epsilon},\nonumber\\
&&z_{\rm in}(z_{\rm out}-z_0)=\frac{J_z-az_{\rm in}}{J_z},
\end{eqnarray}
[which may be readily obtained from \eqref{zinout}, \eqref{eqonz}, and \eqref{auxilir}] one reduces \eqref{xi0} to
\begin{equation}\label{xi1}
\xi(\omega,q,k)=\frac{J_z-az_{\rm in}}{J_z(\omega-\Phi_{\rm bound}(q,k)+i\epsilon)}.
\end{equation}
In order to obtain the final result, one has to find the correct expression for $z_{\rm in}$.

At $b^2-4a^2<0$, \eqref{eqonz} has two solutions
\begin{equation}
z_{\pm}\equiv\frac{-(b+i\epsilon)\pm i\sqrt{4a^2-b^2-2ib\epsilon}}{2a},
\end{equation}
which, following the equality
\begin{equation}
\sqrt{1+ix\epsilon}=x/|x|+\frac{i|x|\epsilon}{2}+o(\epsilon),
\end{equation}
may be reduced to
\begin{equation}\label{zpm1}
z_{\pm}=\frac{1}{2a}\Big[-(b\pm\epsilon)\mp i\Big(\frac{b\sqrt{4a^2-b^2}}{|b|}\pm\epsilon\Big)\Big].
\end{equation}
According to \eqref{zpm1}
\begin{equation}\label{|zpm1|}
|z_{\pm}|^2=1\pm\frac{2b\epsilon}{|b|\sqrt{4a^2-b^2}}+o(\epsilon).
\end{equation}
Following \eqref{zinout} one has
$z_{\rm in}=z_-$ at $b>0$ and $z_{\rm in}=z_+$ at $b<0$. In both the cases, \eqref{|zpm1|} gives
\begin{equation}\label{zin1}
z_{\rm in}|_{\epsilon=0}=\frac{-b+i\sqrt{4a^2-b^2}}{2a}.
\end{equation}
An elementary analysis, with an account for \eqref{omega1}, shows that at $b^2-4a^2>0$ one has
\begin{equation}\label{zin2}
z_{\rm in}|_{\epsilon=0}=\left\{\begin{array}{rcl}
\displaystyle\frac{-b+\sqrt{b^2-4a^2}}{2a},&&\omega\geq\Phi_{\rm up}(q,k)\\
\displaystyle\frac{-b-\sqrt{b^2-4a^2}}{2a},&&\omega\leq\Phi_{\rm down}(q,k)\\
\end{array}
\right..
\end{equation}

Using now the identity
\begin{equation}
b^2-4a^2=D(k,\omega,q),
\end{equation}
which follows from \eqref{omega1} and \eqref{D},
and substituting \eqref{zin1} and \eqref{zin2} into \eqref{xi1}, one readily gets \eqref{k11}.

\item
Substituting the expansions
\begin{widetext}
\begin{eqnarray}
&&\cos{k_{\rm gap}}-\Delta+(1-\Delta\cos{k_{\rm gap}})\cos{k}=(\cos{k_{\rm gap}}-\Delta)(2-(k-k_{\rm gap})^2/2)
+o((k-k_{\rm gap})^2),\nonumber\\
&&\frac{(\Delta\cos{\alpha_{\rm gap}}-\cos{k/2})^2(\Delta-\cos{k/2}\cos{\alpha_{\rm gap}})}{(\Delta-\cos{k/2}\cos{\alpha_{\rm gap}})^2+\cos^2{k/2}\sin^2{\alpha_{\rm gap}}}=
\frac{4\cos^2{k/2}(\Delta-\cos{k_{\rm gap}})^2(\Delta-\cos{k_{\rm gap}}\cos^2{k/2})}{4(\Delta-\cos{k_{\rm gap}}\cos^2{k/2})^2+\sin^2{k}}\nonumber\\
&&=(\Delta-\cos{k_{\rm gap}})^2\frac{(2-k^2/2)[2\Delta-\cos{k_{\rm gap}}(2-k^2/2)]+o(k^2)}
{[2\Delta-\cos{k_{\rm gap}}(2-k^2/2)]^2+k^2+o(k^2)}\nonumber\\
&&=(\Delta-\cos{k_{\rm gap}})^2\frac{4(\Delta-\cos{k_{\rm gap}})+(2\cos{k_{\rm gap}}-\Delta)k^2+o(k^2)}
{4(\Delta-\cos{k_{\rm gap}})^2+(2\Delta\cos{k_{\rm gap}}-1)k^2+o(k^2)}\nonumber\\
&&=(\Delta-\cos{k_{\rm gap}})(1-k^2/4)
+\frac{\Delta\cos{k_{\rm gap}}}{4(\cos{k_{\rm gap}}-\Delta)}k^2+o(k^2),
\qquad\alpha_{\rm gap}\equiv k/2-k_{\rm gap},
\end{eqnarray}
into \eqref{e0e1}, one readily gets \eqref{e(kgap)}. Substituting the expansions
\begin{eqnarray}
&&\cos{k_{\rm gap}}[\cos{k_{\rm max}}-\Delta+(1-\Delta\cos{k_{\rm max}})\cos{k}]=\frac{(\Delta\cos{k_{\rm max}}-1)(k-k_{\rm gap})^2}{2}+o((k-k_{\rm gap})^2),\nonumber\\
&&\Big[\Delta\cos{\Big(\frac{k}{2}-k_{\rm max}\Big)}-\cos{\frac{k}{2}}\Big]^2=\frac{(\Delta\cos{k_{\rm max}}-1)^2(k-\pi)^2}{4}+o((k-\pi)^2),
\end{eqnarray}
into \eqref{e0e1}, one readily gets \eqref{e(kmax)}.
\end{widetext}

\end{itemize}

\begin{widetext}
\begin{figure}[p]
\centering
\includegraphics[width=2.1 \linewidth]{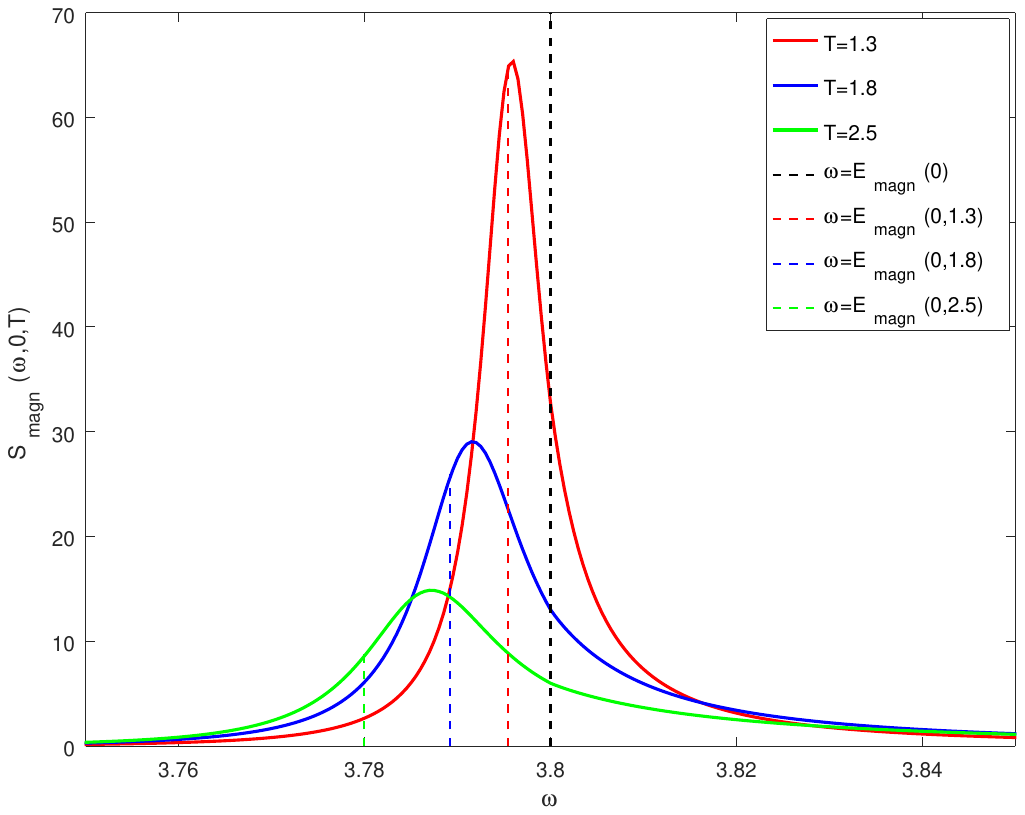}
\caption{Magnon-peak lineshapes $S_{\rm magn}(\omega,0,T)$ at $h=3.0$, $J_z=1$, $J_{\bot}=0.2$ ($E_{\rm gap}=3.8$, $E_{\rm w}=0.4$, $\Delta=5$)
obtained by the formula \eqref{result}.}
\end{figure}
\begin{figure}[p]
\includegraphics[width=2.1\linewidth]{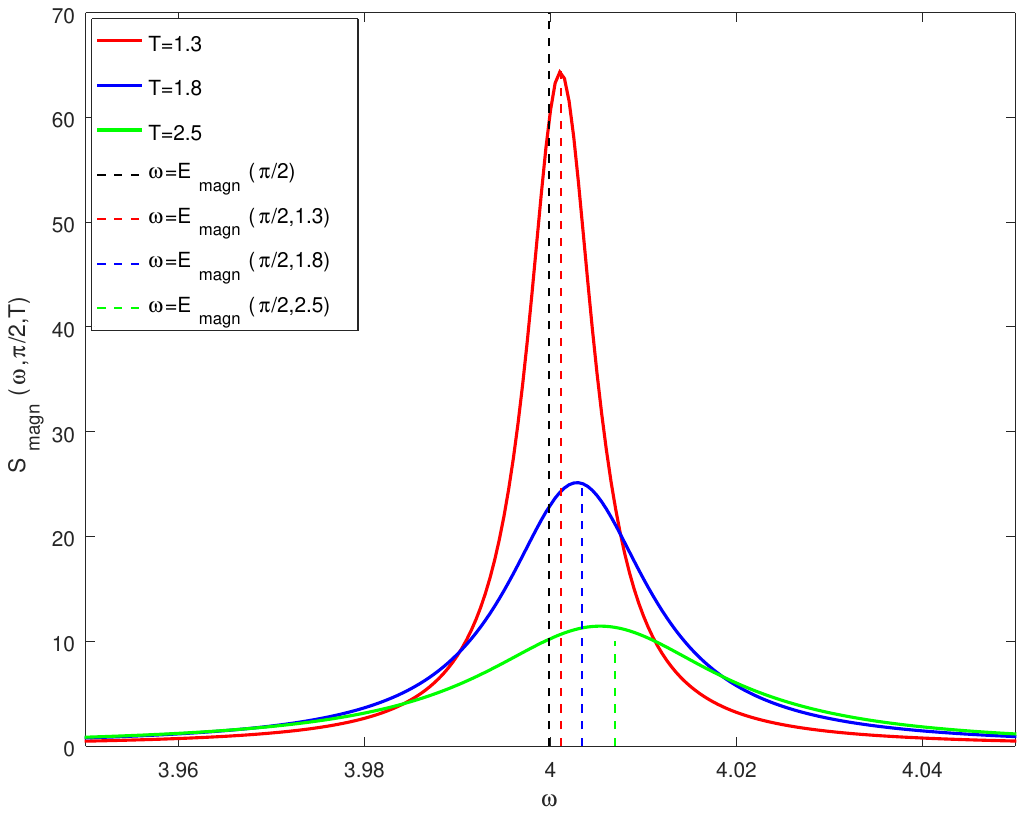}
\caption{Magnon-peak lineshapes $S_{\rm magn}(\omega,\pi/2,T)$ at $h=3.0$, $J_z=1$, $J_{\bot}=0.2$ ($E_{\rm gap}=3.8$, $E_{\rm w}=0.4$, $\Delta=5$)
obtained by the formula \eqref{result}.}
\end{figure}
\begin{figure}[p]
\includegraphics[width=2.1\linewidth]{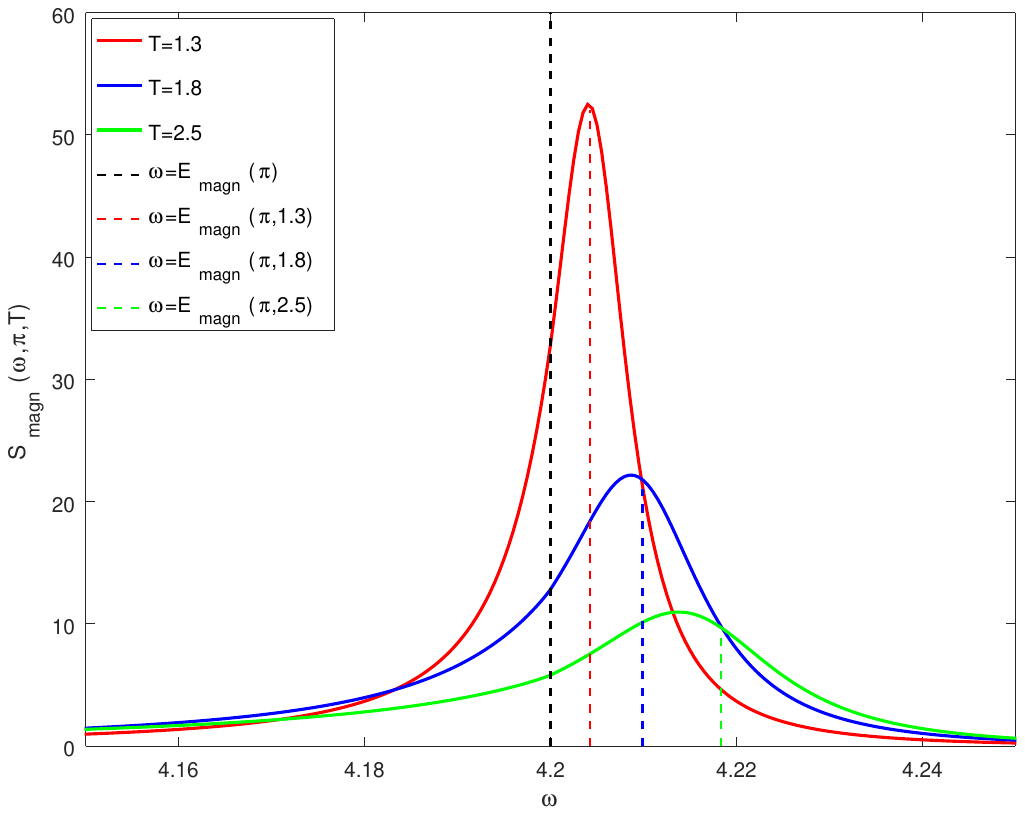}
\caption{Magnon-peak lineshapes $S_{\rm magn}(\omega,\pi,T)$ at $h=3.0$, $J_z=1$, $J_{\bot}=0.2$ ($E_{\rm gap}=3.8$, $E_{\rm w}=0.4$, $\Delta=5$)
obtained by the formula \eqref{result}.}
\end{figure}
\begin{figure}[p]
\includegraphics[width=2.1\linewidth]{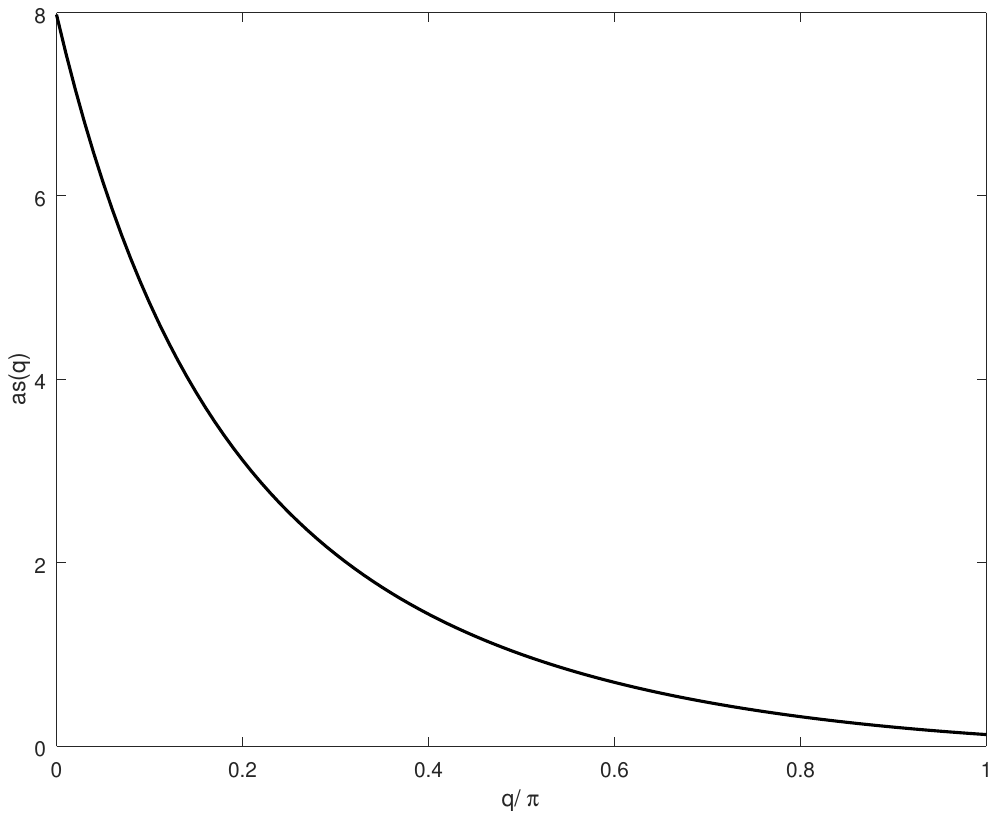}
\caption{Plot of the function ${\tt as}(q)$ defined by \eqref{as}.}
\end{figure}
\begin{figure}[p]
\includegraphics[width=2.1\linewidth]{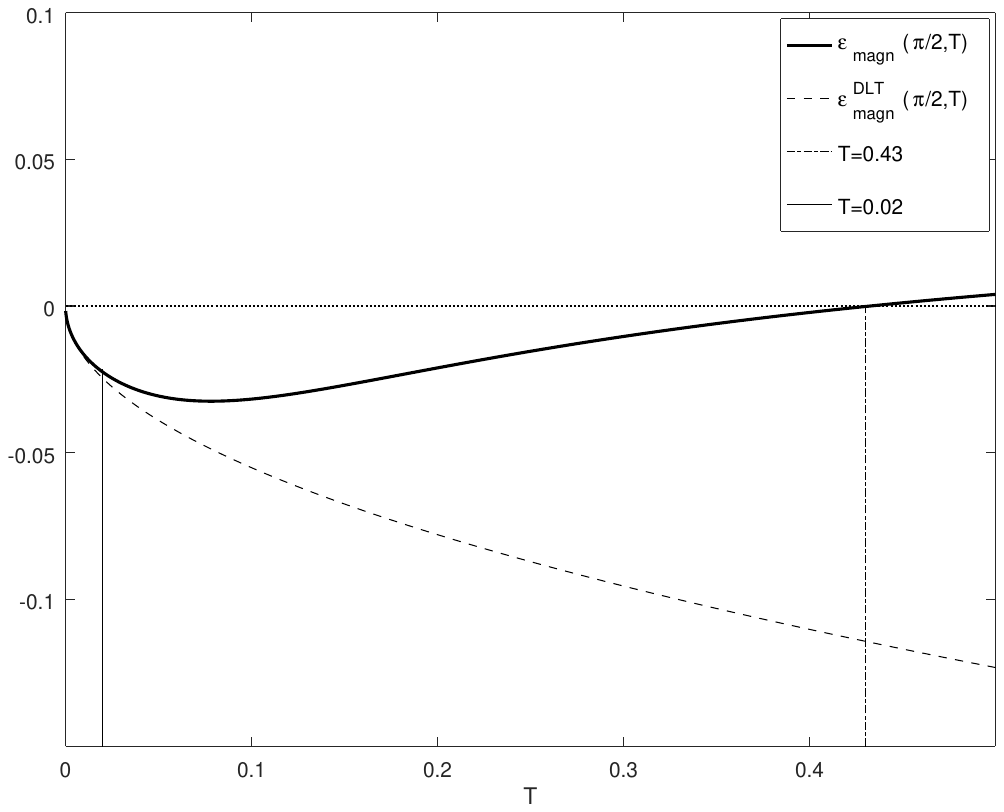}
\caption{Plots for the magnon resonance shift $\varepsilon_{\rm magn}(\pi/2,T)$ obtained by the formulas \eqref{e=e0+e1} and \eqref{e0e1} and its DLT asymptotics $\varepsilon_{\rm magn}^{\rm DLT}(\pi/2,T)$ obtained by \eqref{e0e1DLT}. For $T<0.02$ both the plots are almost identical, however for
$T>0.43$, $\varepsilon_{\rm magn}(\pi/2,T)$ and $\varepsilon_{\rm magn}^{\rm DLT}(\pi/2,T)$ even have different signs.}
\end{figure}
\end{widetext}


\begin{thebibliography}{30}
\bibitem{1} V. E. Korepin, A. G. Izergin, N. M. Bogolubov {\it Quantum Inverse Scattering Method and Correlation Functions} (Cambridge University Press, Cambridge, UK, 1993)
\bibitem{2} C. Trippe, F. ${\rm G\ddot ohmann}$, A. ${\rm Kl\ddot umper}$, Shot-distance thermal correlations in the massive XXZ chain, Eur. Phys. J. Phys. B {\bf73}, 253-264 (2010)
\bibitem{3} M. Brockmann, F. ${\rm G\ddot ohmann}$, M. Karbach, and A. ${\rm Kl\ddot umper}$, On the absorption of microwaves by the one-dimensional spin-1/2 Heisenberg-Ising magnet, Phys. Rev. B {\bf85}, 134438 (2012)
\bibitem{4} F. ${\rm G\ddot ohmann}$, K. K. Kozlowski, and J. Suzuki, High-temperature analysis of the transverse dynamical two-point correlation function of the XX quantum spin chain, J. Math. Phys. {\bf61}, 013301 (2020)
\bibitem{5} F. ${\rm G\ddot ohmann}$, Statistical mechanics of integrable quantum spin systems, SciPost Phys. Lect. Notes, {\bf16} (2020)
\bibitem{6} F. H. L. Essler and R. N. Konik, Application of massive integrable quantum field theories to problems in condenced matter physics, in
{\it From Fields to Strings: Circumnavicating Theoretical Physics} Ian Kogan memorial collection, 684-830 (2005)
\bibitem{7} F. H. L. Essler and R. N. Konik, Finite-temperature lineshapes in gapped spin chains, Phys. Rev. B {\bf 78}, 100403(R) (2008)
\bibitem{8} F. H. L. Essler and R. N. Konik, Finite-temperature dynamical correlations in massive integrable quantum field theories, J. Stat. Mech.: Theor. and Exper., P09018 (2009)
\bibitem{9} W. D. Goetze, U. Karahasanovic, and F. H. L. Essler, Low-temperature dynamical structure factor of the two-leg spin-$\frac{1}{2}$     Heisenberg ladder, Phys. Rev. B {\bf82}, 104417 (2010)
\bibitem{10} P. N. Bibikov, Low-temperature asymptotic of the transverse dynamical structure factor for a magnetically
        polarized $XX$ chain Journ. of Stat. Mech. Theor. and Exper. 073106 (2020)
\bibitem{11} I. A. Zaliznyak and S. H. Lee, Magnetic neutron scattering, in {\it Modern Techniques for Characterizing Magnetic Materials}, edited by Y. Zhu  (Springer, Heidelberg, 2005)
\bibitem{12} Yu. G. Rudou, The Bogoliubov-Tyablikov Green's function method in the quantum theory of magnetism, Theoret. and Math. Phys. {\bf168}, 1318-1329 (2011)
\bibitem{13} N. M. Plakida, The two-time Green's function and the diagram technique, Theoret. and Math. Phys. {\bf168}, 1303-1317 (2011)
\bibitem{14} N. N. Plakida, Dyson equation for Heisenberg ferromagnet, Phys. Lett. A {\bf 43}, 481-482 (1973)
\bibitem{15} Yu. A. Tserkovnikov, Method of solving infinite systems of equations for two-time temperature Green functions, Theoret. and Math. Phys. {\bf49}, 993-1002 (1981)
\bibitem{16} T. Oguchi and I. Ono, Theory of the Ising ferromagnet using the Green function method, Progr. Theor. Phys. {\bf35}, 998-1009 (1966)
\bibitem{17} P. N. Bibikov, Second cluster integral from the spectrum of an infinite $XXZ$ chain, Annals of Phys. {\bf354}, 705-714 (2015)
\bibitem{18} V. V. Val'kov and S. G. Ovchinnikov, Contribution of magnon-magnon interaction to the thermodynamics of anisotropic ferromagnets, Sov. Phys. JETP, {\bf58}, 970-974 (1983)
\bibitem{19} A. L. Chernyshev, M. E. Zhitomirsky, N. Martin, and L.-P. Regnault, Lifetime of gapped excitations in a collinear quantum antiferromagnet, Phys. Rev. Lett. {\bf109} 097201 (2012)
\bibitem{20} I. Poboiko and M. Feigel'man, Spin correlation functions and decay of quasiparticles in XXZ spin chain at $T>0$, Phys. Rev. B {\bf94}, 195420 (2016)
\bibitem{21} A. K. Bera, B. Lake, F. H. S. Essler, L. Vanderstraeten, C. Hubig, U. Schollwock, A. T. M. N. Islam, A. Schneidewind, and D. L. Quintero-Castro, Spinon confinement in a quasi one dimensional anisotropic Heisenberg magnet, Phys. Rev. B {\bf96} 054423 (2017)
\bibitem{22} L. P. Le, A. Keren, M. I. Larkin, G. M. Luke, W. D. Wu, Y. J. Uemura, J. S. Miller, and A. J. Epstein Dynamic spin fluctuations in the molecular ferromagnet (DMeFc)(TCNE), Phys. Rev. B {\bf 65}, 024432 (2001)
\bibitem{23} S. Thota, S. Ghosh, R. Maruthi, D. C. Joshi, R. Medwal, R. S. Rawat, and M. Seehra, Magnetic ground state and exchange interations in the
Ising chain ferromagnet ${\rm CoNb}_2{\rm O}_6$, Phys. Rev. {\bf103}, 064415 (2021)
\bibitem{24} S. Takayoshi, Q. Faure, V. Simonet, B. Grenier, S. Petit, J. Olliver, P. Lejay, and T. Giamarchi, Phase transitions and spin dynamics of the quasi-one dimensional Ising-like antiferromagnet ${\rm BaCo}_2{\rm V}_2{\rm O}_8$ in a longitudinal magnetic field,
Phys. Rev. Research {\bf5}, 023205 (2023)
\bibitem{25} J. E. Gubernatis, N. Kawashima, P. Werner, {\it Quantum Monte Carlo Methods: Algorithms for Lattice Models} (Cambridge University Press, Cambridge, UK 2016)
\bibitem{26} S. Sota and T. Tohyama, Low-temperature density matrix renormalization group using regulared polynomial expansion, Phys. Rev. B {\bf78}
113101 (2008)
\bibitem{27} T. Barthel, U. ${\rm Schollw\ddot ock}$, and S. R. White, Spectral functions in one-dimensional quantum systems at finite temperature
using the density matrix renormalization group, Phys. Rev. B {\bf79} 245101 (2009)
\bibitem{28} M. Nayak and F. Mila, Possibility to detect the bound state of the Heisenberg ferromagnetic chain at intermediate temperature,
Phys. Rev. B {\bf105} 094407 (2022)
\bibitem{29} P. N. Bibikov, Three magnons in an isotropic $S=1$ ferromagnetic chain as an exactly solvable non-integrable system, J. Stat. Mech. {\bf2016}, 033109 (2016)
\bibitem{30} P. N. Bibikov, A three-magnon problem for exactly rung-dimerized spin ladders: From a general outlook to the Bethe ansatz,
J. Phys. A: Math. and Theor. {\bf42}, 315212 (2009)
\end{thebibliography}
\end{document}